\documentclass[twocolumn]{jpsj2}

\usepackage{graphicx}
\usepackage{amsmath}

\begin{document}
\title{
Exact diagonalization study of Mott transition 
in the Hubbard model on an anisotropic triangular lattice
}

\author{
Takashi \textsc{Koretsune}$^{1,2}$, Yukitoshi \textsc{Motome}$^{3}$
 and Akira \textsc{Furusaki}$^{1}$
}
\inst{
$^{1}$Condensed Matter Theory Laboratory,
RIKEN, 2-1 Hirosawa, Wako, Saitama 351-0198\\
$^{2}$Department of Physics, Tokyo Institute of Technology,
2-12-1 Ookayama, Meguro-ku, Tokyo 152-8551\\
$^{3}$Department of Applied Physics, 
University of Tokyo, 7-3-1 Hongo, Bunkyo-ku, Tokyo 113-8656
}

\date{\today}

\abst{
We study Mott transition in 
the two-dimensional Hubbard model on an anisotropic triangular lattice.
We use the Lanczos exact diagonalization of finite-size clusters
up to eighteen sites, and
calculate Drude weight, charge gap, double occupancy and
spin structure factor.
We average these physical quantities
over twisted boundary conditions
in order to reduce finite-size effects.
We find a signature of the Mott transition in
the dependence of the Drude weight and/or charge gap
on the system size.
We also examine the possibility of antiferromagnetic order
from the spin structure factor.
Combining these information, we propose
a ground-state phase diagram
which has a nonmagnetic insulating phase between a metallic phase
and an insulating phase with antiferromagnetic order.
Finally, we compare our results with those reported in the previous
theoretical studies,
and discuss the possibility of an unconventional insulating state.
}

\kword{Mott transition, geometrical frustration, Hubbard model, 
Lanczos exact diagonalization, Drude weight,
averaging over twisted boundary conditions}

\maketitle

\section{Introduction}

Mott transition is a metal-insulator phase transition
driven by strong Coulomb repulsion\cite{mott1949}
and has been a central theme
in physics of strongly correlated electrons.\cite{imada1998}
A Mott insulator is a featureless insulating state
with an odd number of electrons per unit cell
and can be distinguished from a band insulator which has
an even number of electrons per unit cell.
In most cases, however, would-be Mott insulators
show some (translational) symmetry breaking at low temperatures,
resulting in a change in the size of the unit cell.
For example, the Hubbard model and its generalizations
defined on a bipartite lattice
show antiferromagnetic order in their insulating phase
at half filling (one electron per site).
In the narrow definition of a Mott insulator, such insulating states
are not Mott insulators but rather band insulators,
as the unit cell is doubled and contains two electrons.
It is hard to find a true Mott insulator
without any symmetry breaking in nature.

A promising route to a Mott insulator is
to suppress symmetry breaking by introducing
strong geometrical frustration into lattice structure.
A series of layered organic conductors,
$\kappa$-(ET)$_2$X,
have recently attracted much attention as materials that may
realize this route.\cite{kanoda2006,seo2006}
The electronic properties of these compounds are determined by
conducting carriers moving on an anisotropic triangular lattice,
which is geometrically frustrated;
the degree of frustration is controlled by the choice of the anion X.
Furthermore the strength of electron correlation can be easily tuned
by applying pressure.

The $\kappa$-(ET)$_2$X compounds show various intriguing properties.
For example, $\kappa$-(ET)$_2$Cu[N(CN)$_2$]Cl shows a first-order
Mott phase transition without any symmetry breaking at finite temperatures
under pressure.
In the vicinity of the critical end point of the first-order
transition line, the conductivity was found to show
an unconventional critical scaling behavior,\cite{kagawa2005}
which is different from the Ising universality class expected
to describe the Mott criticality in higher (than two)
dimensions.\cite{kotliar2000,limelette2003,castellani1979}
Another compound $\kappa$-(ET)$_2$Cu$_2$(CN)$_3$, whose lattice
structure is close to the isotropic triangular lattice,
has an insulating phase at ambient pressure showing no magnetic order
down to very low temperature, $T=32$mK.\cite{shimizu2003,kurosaki2005,tamura2002}
Some suspect that this insulating phase is a Mott insulator
with a spin-liquid ground state.
When pressure is applied, the Mott insulating state undergoes
a phase transition to a superconducting state.

A minimal theoretical model that contains essential ingredients for
describing the electronic properties of these organic compounds
is a single-band Hubbard model on an anisotropic
triangular lattice at half filling.\cite{kino1996}
It is thus important to understand the Mott
transition and the resulting Mott insulating state in
this and related models.
Indeed there have been several theoretical studies along this line.
In particular, the dynamical mean field theory (DMFT)
has been successfully applied to the study of the Mott transition
in the Hubbard model at finite temperatures.
Among other things it has predicted that the Mott transition at the
critical end point should belong to the Ising universality
class.\cite{kotliar2000}
However, the DMFT is an approach which is justified only in large spatial
dimensions, and it cannot be directly applied to the two-dimensional case.
Some extensions of the DMFT to two dimensions, such as a cellular DMFT,
have also been applied to this model and led to results suggesting
the existence of a first- or second-order transition
without symmetry breaking at finite
temperatures.\cite{onoda2003,parcollet2004,ohashi2006}
However, it is not clear whether spatial correlations are properly
taken into account in this type of approach.
Other theoretical approaches to finite-temperature Mott transitions include
a phenomenological effective theory\cite{imada2005} and a mean-field
theory\cite{misawa2006}, which pointed out an important role played by
a marginal quantum critical point in the unconventional critical
behavior near the critical end point in $\kappa$-(ET)$_2$Cu[N(CN)$_2$]Cl.

The possibilities of the ground state being a non-magnetic insulating state
and a superconducting state have been studied
for the Hubbard model on an anisotropic triangular lattice
using many different approaches such as
the fluctuation exchange approximation,\cite{kino1998,kondo1999}
a U(1) gauge theory,\cite{lee2005}
variational Monte Carlo (VMC) studies\cite{liu2005,watanabe2006} and
the cellular DMFT.\cite{kyung2006}
Related models were also discussed, for example, in
the variational study of the Heisenberg spin model 
with ring exchange coupling\cite{motrunich2005} and
the resonating-valence-bond mean field theory for the Hubbard-Heisenberg
model.\cite{gan2005,powell2005}
These studies, however, often rely on some uncontrolled approximations
whose validity is uncertain
when both the strong correlation and the geometrical frustration
are important.

For the half-filled Hubbard model with geometrical frustration,
choices are quite limited
of numerical approaches that can both treat correlations
in real space and work at zero temperature (or low temperatures).
For example, the DMFT and its cluster extensions do not
fully treat the correlations in the real space, as we have mentioned.
Monte Carlo simulations are susceptible to the infamous
negative sign problem, and work only at high temperatures.
At this stage, available numerical methods that satisfy the required conditions are
the path integral renormalization group (PIRG)
method\cite{imada2000,kashima2001}
and the exact diagonalization method.

Imada and his coworkers have developed the PIRG method and applied it
to study the Hubbard model on two-dimensional frustrated lattices.
In the ground-state phase diagram of the half-filled Hubbard model
on an anisotropic triangular lattice, they found
a narrow region of a non-magnetic insulating phase
between a paramagnetic metal and an antiferromagnetic insulator,
if the frustration is substantial.\cite{morita2002}
In the non-magnetic insulating state, no obvious symmetry breaking
pattern was observed, and energy spectrum has highly degenerate
low-lying excitations
as in the Fermi liquid, which could be attributed to the presence
of a spinon Fermi surface.\cite{motrunich2005,lee2005}
The transition from the metallic phase to the non-magnetic insulator
appears to be first order although the discontinuity becomes weaker
for strong frustration.
In order to confirm and better understand these interesting behaviors,
a complementary study using another unbiased technique is needed.

To this end,
we use the exact diagonalization method
to study the ground-state properties of 
the two-dimensional Hubbard model on the anisotropic
triangular lattice at half filling.
The method treats correlation effects in an unbiased way
with high precision,
although the calculation is limited to small systems.
We examine dependence on boundary conditions (BCs) for finite-size clusters,
and demonstrate that averaging over the twisted BCs
is an efficient tool to reduce the finite-size effects.
We calculate the Drude weight, which was not calculated in the previous
works, to monitor the metal-insulator transition directly.
We also compute the charge gap, the double occupancy and
the spin structure factor 
to elucidate the phase diagram and the nature of the transitions.
The results are discussed in comparison with those in the previous works.

This paper is organized as follows.
In \S2, we define the model and finite-size clusters
to be studied numerically.
The physical quantities signalling the Mott transition
are also introduced.
In \S3, we discuss their dependences on
BCs in the finite-size clusters,
and demonstrate that averaging over the twisted BCs
leads to substantial reduction of the finite-size effects.
In \S4, we present numerical results on the ground-state properties.
The numerical results and their implications to
the phase diagram are discussed in \S5.
We conclude with a brief summary in \S6.

\section{Model and Method}

\begin{figure}
\begin{center}
\includegraphics[scale=0.6]{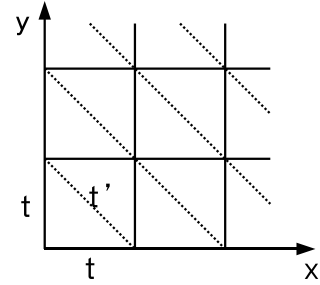}
\caption{Anisotropic triangular lattice and the hopping integrals in our model (\ref{eq:H}).}
\label{fig:lattice}
\end{center}
\end{figure}

We study the ground-state properties of
the Hubbard model on an anisotropic triangular lattice at half filling.
Its Hamiltonian is given by
\begin{equation}
H = -\sum_{\langle ij \rangle \sigma}
t_{ij}( \chi_{ij} c_{i \sigma}^{\dagger} c_{j \sigma}^{} + \mathrm{H.c.} )
+ U \sum_i n_{i \uparrow} n_{i \downarrow},
\label{eq:H}
\end{equation}
where $c_{i \sigma}$ $(c_{i \sigma}^{\dagger})$ is the annihilation (creation)
operator for electrons of spin $\sigma$ on the site $i$,
and $n_{i \sigma} = c_{i \sigma}^{\dagger} c_{i \sigma}^{}$ is
the number operator.
The model is defined on the anisotropic triangular lattice
with the hopping integrals $t_{ij}$
as shown in Fig.\ \ref{fig:lattice}.
We take $t = 1$ as the energy unit and
fix the electron density at half filling, one electron per site
on average.

\begin{figure}
\begin{center}
\includegraphics[scale=0.35]{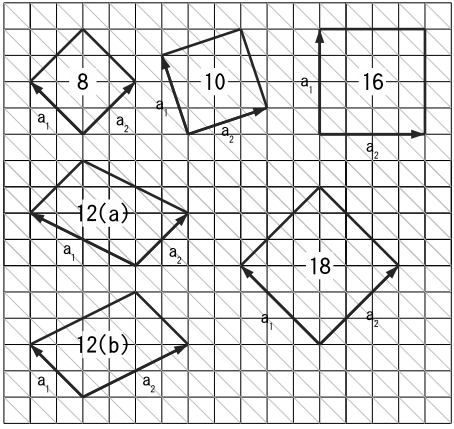}
\caption{Finite-size clusters used in our study.
The number indicates the size of each cluster, $N$.
There are two different geometries for the 12-site clusters.
All the clusters are compatible with antiferromagnetic
N\'eel order while only the 12(a)- and 18-site clusters are
compatible with 3-sublattice order.
}

\label{fig:cluster}
\end{center}
\end{figure}

We consider finite-size clusters on the triangular lattice
shown in Fig.\ \ref{fig:cluster}.
To impose a twisted BC for each cluster,
we have introduced the phase factor,
\begin{equation}
\chi_{ij} = \exp(\mathrm{i} \mib{\phi} \cdot \mib{r}_{ij}^{}),
\end{equation}
where $\mib{r}_{ij}$ is the vector connecting from the site $i$
to the site $j$, and
$\mib{\phi}
=\phi_1\mib{b}_1+\phi_2\mib{b}_2$ in terms of
the reciprocal vectors $\mib{b}_1$ and $\mib{b}_2$,
which satisfy the orthonormal relations
$\mib{a}_i\cdot\mib{b}_j=\delta_{i,j}$
with the vectors $\mib{a}_1$ and $\mib{a}_2$
defining the cluster (Fig.\ \ref{fig:cluster}).
The ``flux'' $\mib{\phi}=(\phi_1,\phi_2)$
dictates the BC
along the direction parallel to the vectors $\mib{a}_1$ and $\mib{a}_2$.
For example, $\mib{\phi} = (0, 0)$ represents periodic BCs in both directions,
and $\mib{\phi} = (0, \pi)$ periodic BC in the $\mib{a}_1$ direction
and anti-periodic BC in the $\mib{a}_2$ direction, respectively.
Any nonvanishing flux $\mib{\phi}$ leads to a twisted BC
($-\pi\le\phi_i\le\pi$, $i=1,2$).

We apply the exact diagonalization method based on the Lanczos technique
to finite-size clusters up to 
18 sites shown in Fig.\ \ref{fig:cluster}.
Since the model has translational symmetry,
each energy eigenstate is classified according to
its total momentum $\mib{k}$.
For a given $\mib\phi$, we numerically obtain a ground-state wave
function of a finite-size cluster for several values of $U$ and $t'$.
To save the CPU time, for 16-site and 18-site clusters, 
lowest-energy states are computed
at some relevant momenta only, such as
$\mib{k} = (0,0), (\pi,\pi), (\pi,0)$, $(0,\pi)$,
and the momentum of the ground state at $U=0$.
These wave functions on finite-size clusters depend on $\mib\phi$.
As we discuss in the next section, the $\mib\phi$-dependence can
be used for reducing finite-size effects.
Here let us assume for the moment that $\mib\phi$ is
fixed at a certain value.

To study the metal-insulator transition at zero temperature,
we compute the Drude weight and the charge gap.
The Drude weight is given by\cite{kohn1964,scalapino1992,scalapino1993}
\begin{align}
\frac{D_\mu}{2 \pi e^2} &=
\frac{1}{2N} \langle 0;\mib{\phi} | F_{\mu\mu} | 0;\mib{\phi} \rangle
+ \frac{1}{N} \sum_{n\neq0}
  \frac{ |\langle n;\mib{\phi} | J_\mu | 0;\mib{\phi} \rangle|^2 }
       {E_0(\mib{\phi}) - E_n(\mib{\phi})}
\nonumber\\
&
= \frac{1}{2N} \frac{\partial^2 E_0(\mib{\phi}) }{\partial \phi_\mu^2}
\label{eq:drude}
\end{align}
with
$J_\mu = \partial H/\partial \phi_\mu$,
$F_{\mu\nu} = \partial^2 H/\partial \phi_\mu
 \partial \phi_\nu$ ($\mu, \nu = x, y$) and
$N$ being the number of sites. 
Here, $E_0(\mib{\phi})$ and $E_n(\mib{\phi})$ are
the energy eigenvalues of the ground state
$|0;\mib{\phi}\rangle$ and 
of the $n$-th excited state $|n;\mib{\phi}\rangle$, respectively.
We use the second equation in eq.\ (\ref{eq:drude})
to calculate the Drude weight, i.e.,
the second derivative of the ground-state energy
with respect to the flux $\mib\phi$. 
We define the dimensionless Drude weight
$D = (D_x + D_y)/(4 \pi e^2)$.\cite{note2} 

The charge gap is defined as a change in the chemical potential
when electrons are added to or subtracted from the system,
\begin{equation}
\Delta_{\rm c} = \mu^+ - \mu^-,
\label{eq:Delta_c}
\end{equation}
where
\begin{align}
\mu^+ &= \frac{1}{2}[E_0(N+2,\mib{\phi}) - E_0(N,\mib{\phi})],
\label{eq:mu+}\\
\mu^- &= \frac{1}{2}[E_0(N,\mib{\phi}) - E_0(N-2,\mib{\phi})].
\label{eq:mu-}
\end{align}
Here $E_0(N_e,\mib{\phi})$ is the ground-state energy with $N_e$ electrons, and
the number of electrons is changed by two to conserve 
the $z$ component of the total spin.

We also calculate the double occupancy,
\begin{equation}
d_\mathrm{occ} = \frac{1}{N} \sum_{i}
 \langle n_{i \uparrow} n_{i \downarrow} \rangle_{\mib{\phi}},
\end{equation}
which is believed to be related with the order parameter of
the Mott transition.
Here and henceforth $\langle O \rangle_{\mib{\phi}}$ is
understood as the average value of the operator $O$
in the lowest-energy state for the given $\mib\phi$,
$\langle O \rangle_{\mib{\phi}}=
 \langle0;\mib{\phi}| O |0;\mib{\phi}\rangle$.

To study the magnetic properties,
we calculate the spin structure factor,
\begin{equation}
S(\mib{q}) = \frac{1}{N} \sum_{ij} \langle S_i^z S_j^z \rangle_{\mib{\phi}}
e^{-\mathrm{i} \mib{q} \cdot \mib{r}_{ij}^{}}.
\label{eq:sq}
\end{equation}
We note that $D$, $\Delta_\mathrm{c}$, $d_\mathrm{occ}$,
and $S(\mib{q})$ are
functions of $\mib\phi$.

\section{Averaging over twisted boundary conditions}

In order to extract intrinsic properties of a bulk system,
we need to extrapolate numerical results of finite-size
clusters to the thermodynamic limit $N \to \infty$.
In practice, for small clusters that can be dealt with the
exact diagonalization technique, the numerical results
have large finite-size effects and strong dependence
on the BC or $\mib\phi$.
Understanding this $\mib\phi$-dependence is an essential step
towards systematic analysis of finite-size effects and extrapolation
to the thermodynamic limit.\cite{fye1991}

\begin{figure}
\includegraphics[scale=0.6]{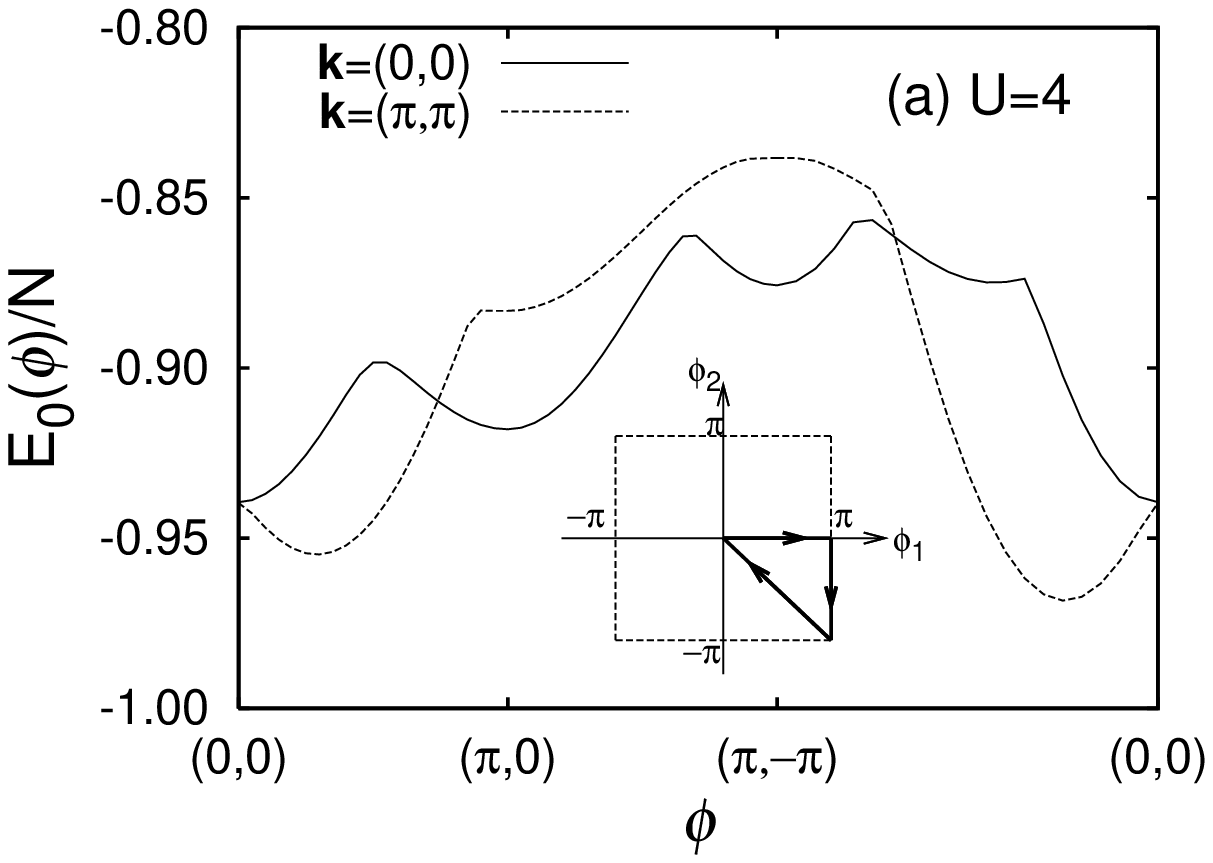}
\includegraphics[scale=0.6]{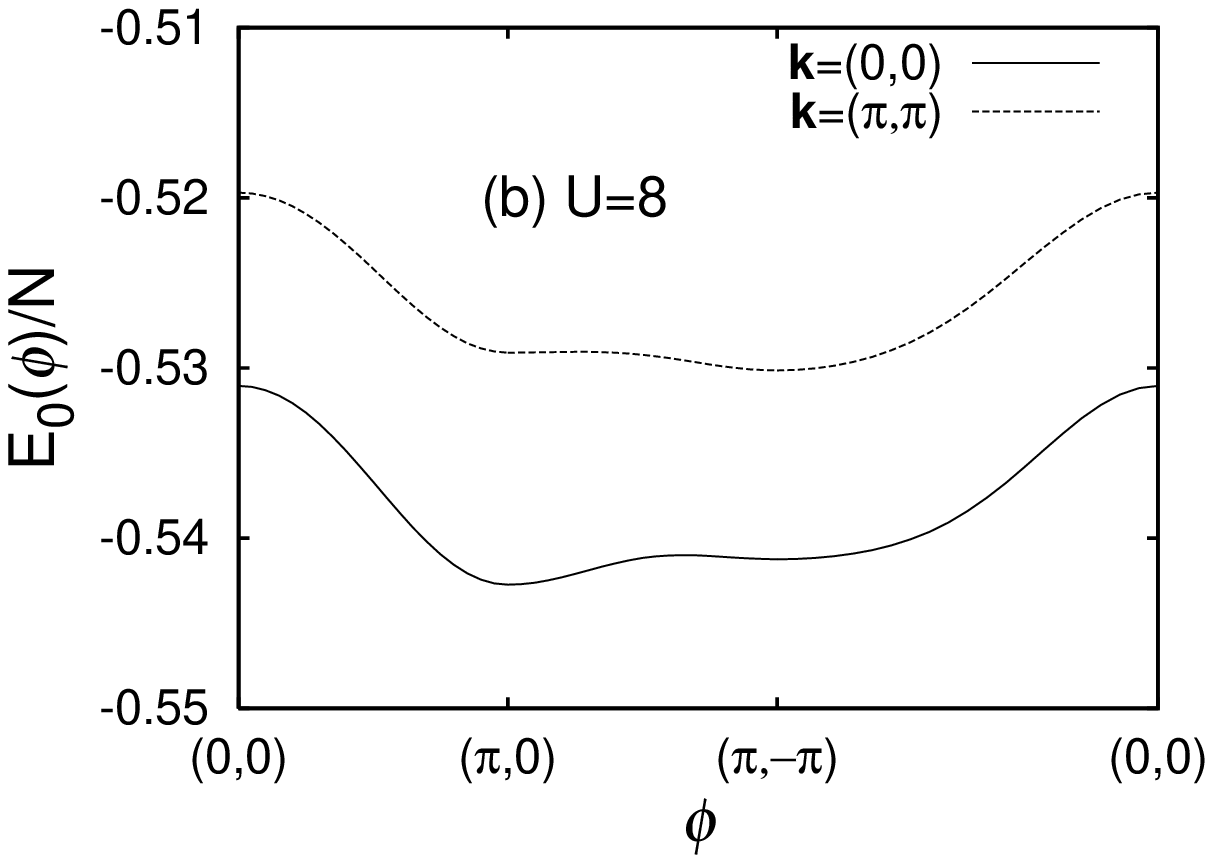}
\caption{
Boundary condition dependence of the lowest energies with
$\mib{k}=(0,0)$ and
$(\pi,\pi)$ at (a) $U=4$ and (b) $U=8$ for $t'=0.5$ on 16-site cluster.
The inset in (a) shows the parametrization of $\mib{\phi}$.
}
\label{fig:bcdep}
\end{figure}

To illustrate the problem we are facing,
we first show $\mib\phi$-dependence of the lowest energies
of $\mib{k}=(0,0)$ and $(\pi,\pi)$
at $U=4$ and $8$ for $t'=0.5$ on the 16-site cluster
in Fig.\ \ref{fig:bcdep}.
We clearly see that at $U=4$ [Fig.\ \ref{fig:bcdep}(a)],
the ground-state momentum changes from $\mib{k}=(0,0)$
to $(\pi,\pi)$ and vice versa as $\mib\phi$ varies.
This implies that the states with different momenta become degenerate
ground states
in the thermodynamic limit, which is a typical feature of
a Fermi liquid (metal).
On the other hand, as shown in Fig.\ \ref{fig:bcdep}(b),
there is no momentum change in the ground state at $U=8$,
indicating the insulating behavior in the thermodynamic limit
where a ground state is singled out with opening of a gap.\cite{note}
A similar behavior is seen already in the smallest 8-site cluster.

Typically, there are three different ways to treat the BCs
in the finite-size analysis:
(i) fixing the BC,\cite{dagotto1992} i.e., fixing $\mib\phi$
at a priori chosen value,
for example, at $\mib\phi=(0,0)$;
(ii) optimizing $\mib\phi$ for each cluster size and parameter set to
minimize the ground-state energy,\cite{nakano1999,nakano2007} and;
(iii) averaging over all the BCs
computed physical quantities.\cite{poilblanc1991,gros1992}
The results obtained using these methods will,
of course, converge to the same result
in the thermodynamic limit, but can have quite different
system-size dependences.

The first approach (i) has ambiguity
in choosing an appropriate BC for analyzing the size dependence,
as finite-size effects appear very differently with different BCs.
For example, when $U$ is changed in the 16-site cluster, a level crossing
(or first-order transition) occurs for $\mib\phi=(0,0)$,
while no transition occurs for $\mib\phi=(\pi,0)$.
Thus, there is no obvious choice for a fixed and optimum BC,
and we do not take this approach.

\begin{figure}
\includegraphics[scale=0.6]{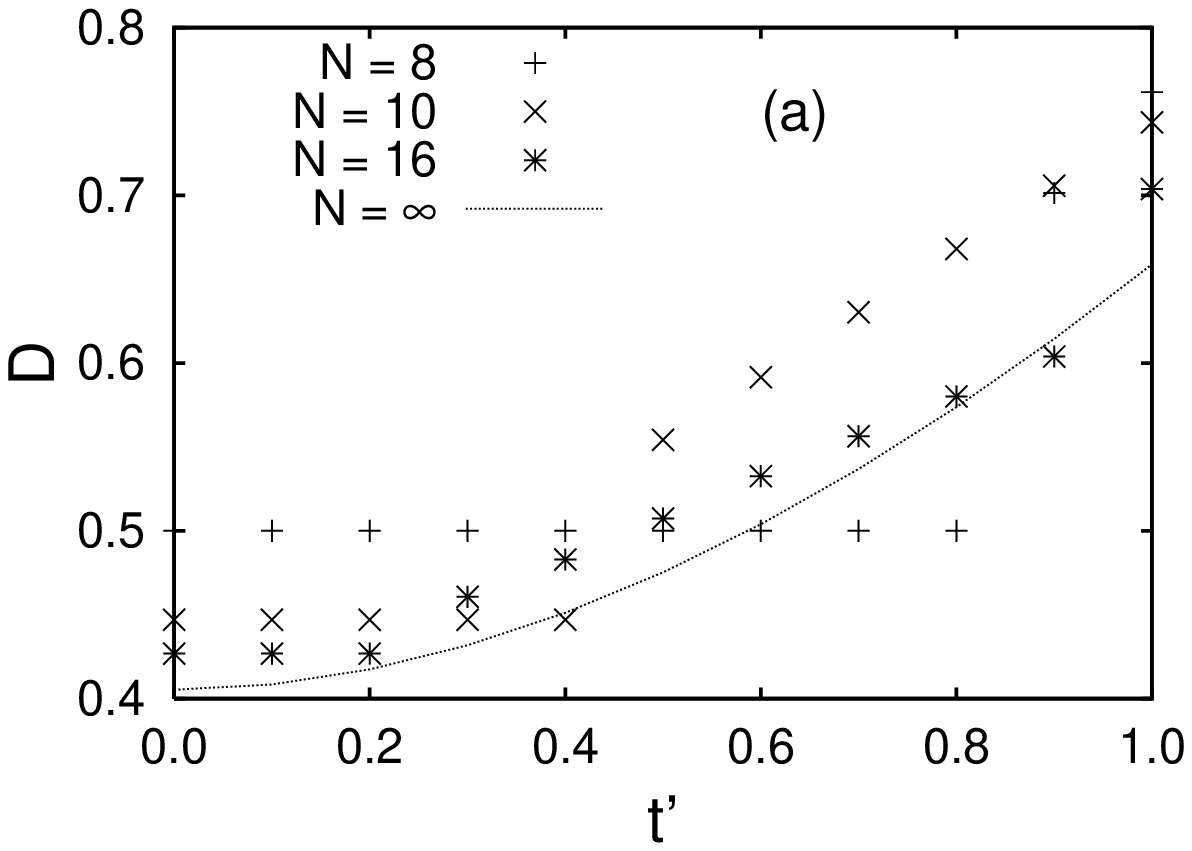}
\includegraphics[scale=0.6]{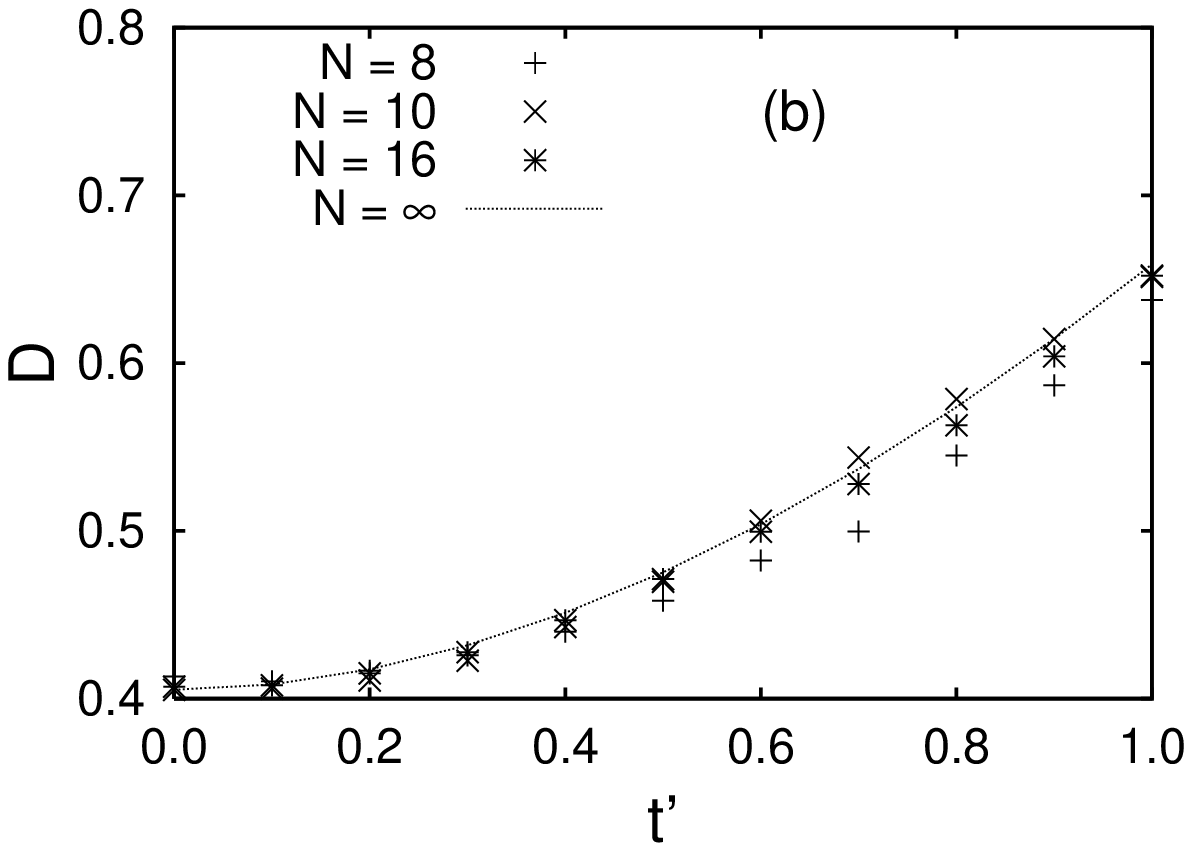}
\caption{
$t'$ dependence of the Drude weight at $U=0$ obtained by
(a) searching BC which minimizes the ground-state energy and
(b) averaging over BC.
}
\label{fig:u0}
\end{figure}

The second approach (ii)
also has a problem. 
For example, at $U=0$
the ground-state momentum changes at a finite value of $t'$
that depends on the system size;
see Fig.\ \ref{fig:u0}(a) below.
Thus, 
it is difficult to obtain a smooth, systematic system-size dependence
with this method.

Compared with these two, 
the approach (iii), averaging over the BCs, does not have
such difficulties and shows a well-behaved system-size dependence 
as we demonstrate below.
In this approach, average is taken
over the BCs for each cluster size,\cite{poilblanc1991,gros1992}
\begin{equation}
\langle O \rangle_{\rm BC} = 
\frac{1}{\mathcal{A}}
\int d{\boldsymbol \phi} \; \langle O \rangle_{\mib{\phi}},
\end{equation}
where $\mathcal{A}=(2\pi)^2|\mib{b}_1\times\mib{b}_2|$, and
$O$ is a physical quantity (such as the Drude weight $D$)
introduced in the previous section.
In the numerical calculations, 
the integral over the Brillouin zone is approximated
by the summation over grid points.
Error bars are estimated as $\sigma/\sqrt{N_\phi}$, where
$\sigma$ is the standard deviation and
$N_\phi$ is the number of grid points.
We take sufficiently large $N_\phi$ for 8- and 10-site clusters.
For 16-site cluster, we choose $N_\phi = 32\sim128$ 
to obtain the precisions required.
For 18-site cluster, we take $N_\phi = 16$.

Let us demonstrate the efficiency of the averaging procedure
in comparison with the method (ii).
Figure \ref{fig:u0} shows $t'$ dependence of the Drude weight at $U=0$
obtained by the two methods (ii) and (iii).
In Fig.\ \ref{fig:u0}(a) [the method (ii)],
the Drude weight of each cluster size shows
a sudden change (kink) in the $t'$ dependence,
which corresponds to the change of the ground-state momentum.
The kink appears at different values of $t'$
for different system sizes, which complicates analysis of
the system-size dependence in the method (ii).
Furthermore, the results of finite $N$ show large
deviations of about 10\% from the Drude weight in the thermodynamic limit.
By contrast, Fig.\ \ref{fig:u0}(b) shows that
the Drude weight averaged over the BCs [the method (iii)]
at finite $N$
is a smooth function of $t'$ and is very close to that in the
thermodynamic limit;
The differences between the data points of $N=16$ and $N=\infty$
are less than 1\%.

This good convergence, seen already at small $N$,
can be understood as follows.
At $U=0$ the Drude weight can be written in the thermodynamic limit as
\begin{equation}
D = \frac12 \int \frac{d\mib{k}}{(2\pi)^2}\, g(\mib{k})
    \Theta(\epsilon_{\rm F}-\epsilon_{\mib{k}}),
\label{D in the thermodynamic limit}
\end{equation}
where $\Theta(x)$ is the Heaviside step function,
$\epsilon_{\rm F}$ the Fermi energy, and
\begin{equation}
g(\mib{k}) = 2t (\cos k_x + \cos k_y) + 4t' \cos(k_x-k_y).
\end{equation}
The single-particle energy $\epsilon_{\mib{k}}$ is given by
\begin{equation}
\epsilon_{\mib{k}} = -2t(\cos k_x + \cos k_y) - 2t' \cos(k_x - k_y).
\end{equation}
For a finite-size cluster the averaged Drude weight 
is calculated as
\begin{equation}
\langle D \rangle_{\rm BC} =
\frac{1}{2\mathcal{A}} \int d\mib{\phi} 
\frac{1}{N}\sum_{\mib{k}}{}' \,
g(\mib{k}+\mib{\phi}),
\label{<D>_BC}
\end{equation}
where the primed sum denotes summation over
$N$ lowest-energy single-particle states
of energy $\epsilon_{\mib{k}+\mib{\phi}}^{}$.
We note that the number of single-particle states inside the
Fermi surface determined by $\epsilon_{\mib{k}}=\epsilon_{\rm F}$
is not always equal to $N$ in finite-size clusters.
Up to this finite-size effect eq.\ (\ref{<D>_BC}) is
equal to eq.\ (\ref{D in the thermodynamic limit}).
This indicates that, in the non-interacting case,
the thermodynamic limit can be reached by averaging over the BCs
even in small clusters.\cite{bonca2003}
Note that the method (iii) can distinguish between a band metal and
a band insulator.

\begin{figure}
\includegraphics[scale=0.6]{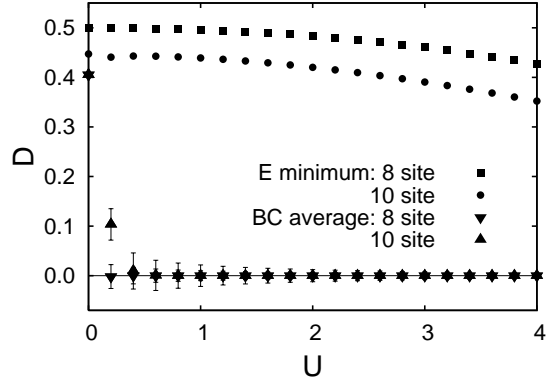}
\caption{
$U$ dependence of the Drude weight at $t'=0$ obtained by
two different methods as in Fig.\ \ref{fig:u0}.
}
\label{fig:tp0}
\end{figure}

Next, Fig.\ \ref{fig:tp0} compares $U$ dependences of
the Drude weight at $t'=0$ obtained with
the two methods (ii) and (iii).
This $t'=0$ case corresponds to the Hubbard model on the square lattice,
in which the metal-insulator transition is known to take place
at $U_{\rm c} = 0$ because of the perfect nesting.
As shown in Fig.\ \ref{fig:tp0}, however,
the Drude weight computed with BCs minimizing the energy
[the method (ii)]
remains finite even when $U\gg t$.
A similar behavior was observed in the one-dimensional case.\cite{fye1991}
In contrast to this,
the Drude weight averaged over all the BCs [the method (iii)]
immediately vanishes for $U > 0$,
reproducing the expected metal-insulator transition,
except for $U\ll t$
where the averaged Drude weight will
converge slowly to zero with $N_\phi \rightarrow \infty$.

The reason why the expected behavior is reproduced
even with the small size clusters
is understood as follows.
Let us consider the $L\times L$ clusters for simplicity.
At $U=0$ the ground-state momentum is $\mib{k}=(0,0)$
for any $\mib\phi$ because of the perfect nesting.
In the $\phi_1$-$\phi_2$ plane
the ground-state energy has cusp singularities along the lines,
$\phi_1 = \phi_2$ and $\phi_1 = -\phi_2$,
where the ground states are degenerate.
As we discussed above,
the Drude weight averaged over the BCs reproduces well
the one in the $N\to\infty$ limit,
when the cusp singularities are avoided 
in the average summation.
A small but finite $U$ lifts the degeneracy, and
the cusp lines disappear.
Then the ground-state energy becomes an analytic function of $\mib\phi$,
and the average Drude weight
vanishes by definition,
for instance,
$
\langle D_x \rangle_{\rm BC} \propto 
\int d{\boldsymbol \phi} \; 
\partial^2 E_0 / \partial \phi_x^2 = 0.
$
This leads to the metal-insulator transition at $U\ll t$ expected 
for this case with the perfect nesting.

From the results shown above, we conclude that
the averaging over the twisted BCs
is most efficient among the three methods to calculate $D$.
We have applied this method to calculate
the double occupancy and the spin structure factor,
and confirmed its efficiency for these quantities as well.
In the following section, we present numerical results
of the Drude weight, the double occupancy and
the spin structure factor, which are obtained with this method.
(We will omit the brackets $\langle \cdots \rangle_{\rm BC}$.)

It turned out, however, that the method (iii) is not suitable
for calculating the charge gap.
To illustrate its drawback, let us consider a band insulator at $U=0$.
In this case, the charge gap is nothing but a band gap.
If we calculate the charge gap (\ref{eq:Delta_c}) 
from the chemical potentials $\mu_\pm$ averaged over BC, 
the gap is 
overestimated, since 
the averaged chemical potentials $\mu^+$ and $\mu^-$ correspond to
the average energy at the bottom 
of a conduction band and 
that at the top of a valence band, respectively.
To reproduce the band gap in the thermodynamic limit,
we should instead search the minimum energy of the conduction band
and the maximum energy of the valence band. 
From this consideration,
we use the definition 
\begin{equation}
\Delta_{\rm c} = \max \left\{ \mu_{\rm min}^{+} - \mu_{\rm max}^{-} 
\, , \; 0 \right\},
\label{eq:chargegap}
\end{equation}
where
\begin{align}
\mu_{\rm min}^{+} &=
\min_{\boldsymbol \phi} \frac{E_0(N+2, {\boldsymbol \phi})
 - E_0(N, {\boldsymbol \phi})}{2},\\
\mu_{\rm max}^{-} &=
\max_{\boldsymbol \phi} \frac{E_0(N, {\boldsymbol \phi})
 - E_0(N-2, {\boldsymbol \phi})}{2}.
\end{align}
With this definition,
finite-size effects are completely removed for the band gap.
Furthermore we obtain $\Delta_{\rm c} = 0$
at the noninteracting case $U=0$ as expected for a metallic state.
These simple arguments suggest that
the definition of eq.\ (\ref{eq:chargegap}) should be
most effective for extracting the charge gap
in the thermodynamic limit. 
We employ this definition in \S4.2.

\section{Results}
In this section,
we present our numerical results for $D$, $\Delta_\mathrm{c}$,
$d_\mathrm{occ}$ and
$S(\mib{q})$ computed for a wide parameter range
of $t'$ and $U$
with the method described in the previous sections.

\subsection{Drude weight}
\begin{figure}
\includegraphics[scale=0.6]{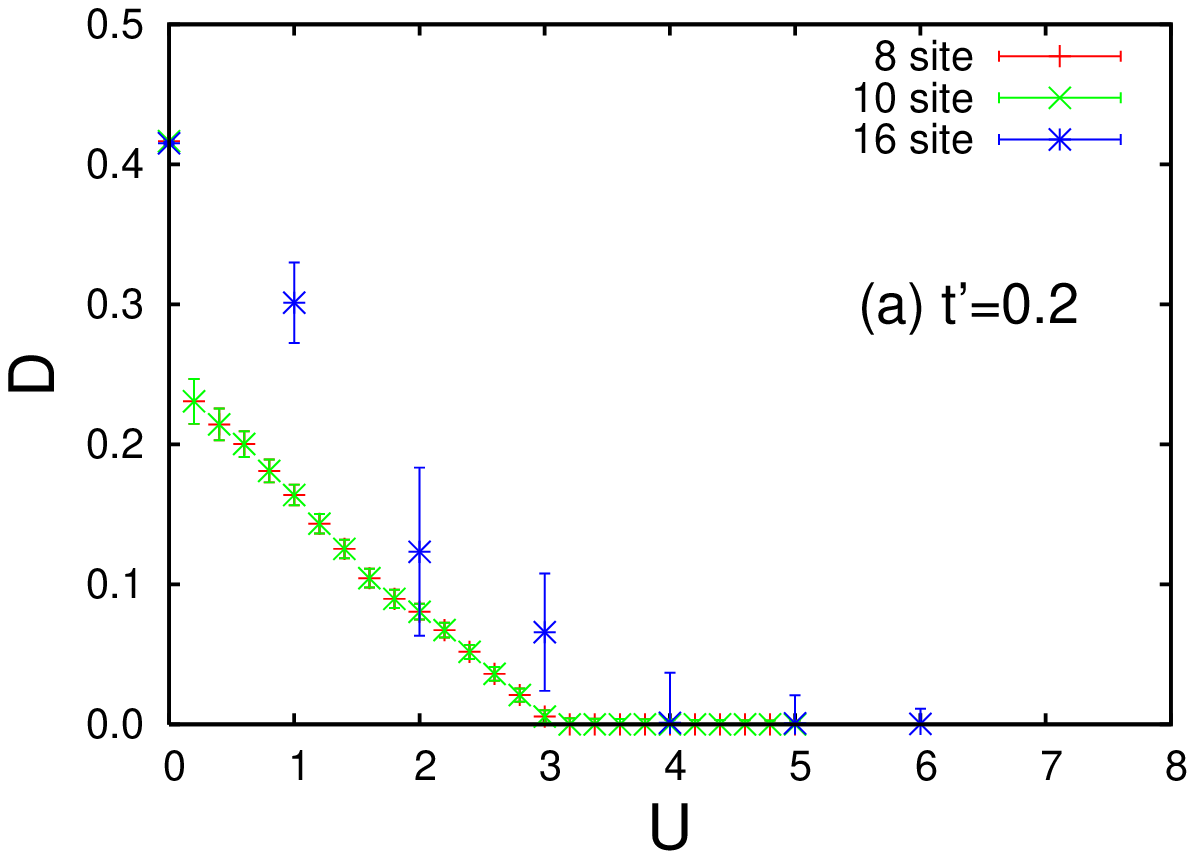}
\includegraphics[scale=0.6]{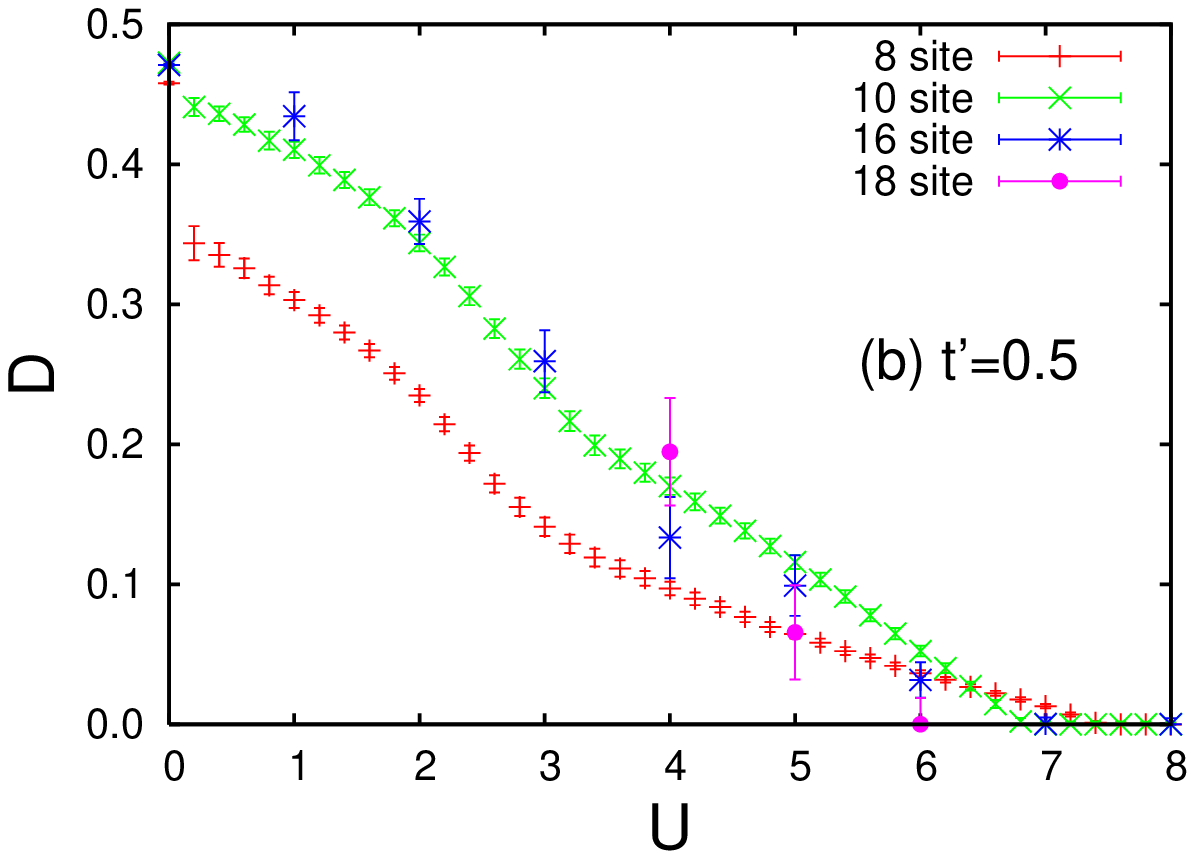}
\includegraphics[scale=0.6]{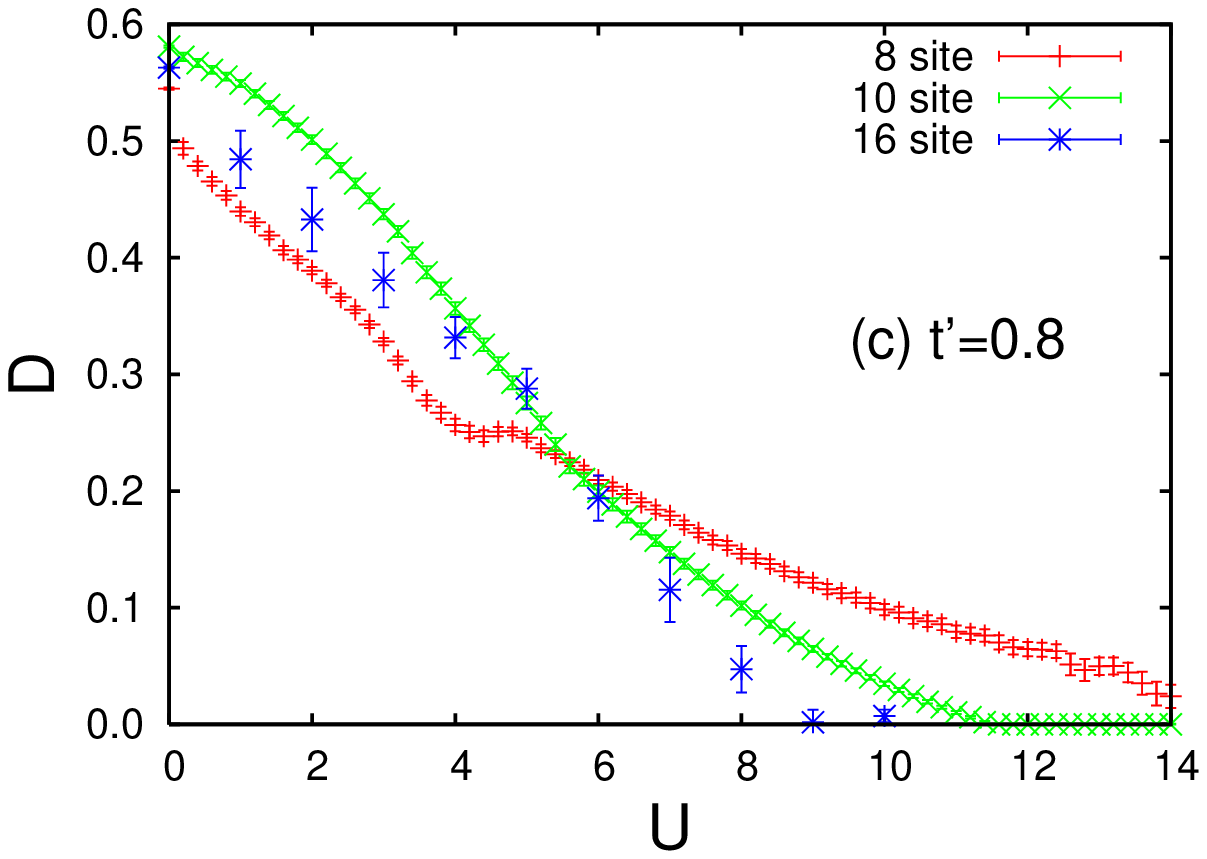}
\caption{
Drude weight at (a) $t'=0.2$, (b) $0.5$ and (c) $0.8$ 
for 8-, 10- and 16-site clusters.
At $t'=0.5$, data for 18-site cluster are also plotted.}
\label{fig:drude}
\end{figure}

Figure \ref{fig:drude} shows the Drude weight at $t'=0.2$, $0.5$ and $0.8$.
For each cluster size, the Drude weight decreases with increasing $U$
and eventually vanishes.
We can thus define the critical value $U_\mathrm{c}^D$
above which the Drude weight vanishes.
The critical value $U_\mathrm{c}^D$ is expected to represent
the metal-insulator transition point.
At $t'=0.2$ and $0.5$, the system-size dependence of $U_\mathrm{c}^D$
is small, and it is reasonable to estimate
$U_\mathrm{c}^D = 3 \sim 4$ for $t'=0.2$
and $U_\mathrm{c}^D = 5 \sim 7$ for $t'=0.5$, respectively.
However, $U_\mathrm{c}^D$ at $t'=0.8$ shows a substantial
system-size dependence.
Considering the fact that the Drude weight at $U=6$ is almost
independent of the system size,
we can roughly estimate $U_{\rm c}^{D} = 6\sim8$.

Let us remark on the discontinuity of the Drude weight
at $U=0$ seen in the numerical results for small clusters.
This discontinuity is an artifact of small system size and
can be traced back to the cusp singularity at $t'=0$
mentioned in \S3.
In small clusters the number of discrete momenta is small, and 
the region represented by $\mib{k}=(0,0)$ is relatively large.
Hence, the effect of the collapse of the cusp lines becomes
pronounced and leads to
the discontinuity of the Drude weight even at finite $t'$.
In fact, the jump in the Drude weight is larger for smaller clusters
and for smaller $t'$ as shown in Fig.\ \ref{fig:drude}.

\subsection{Charge gap}
\begin{figure}
\includegraphics[scale=0.6]{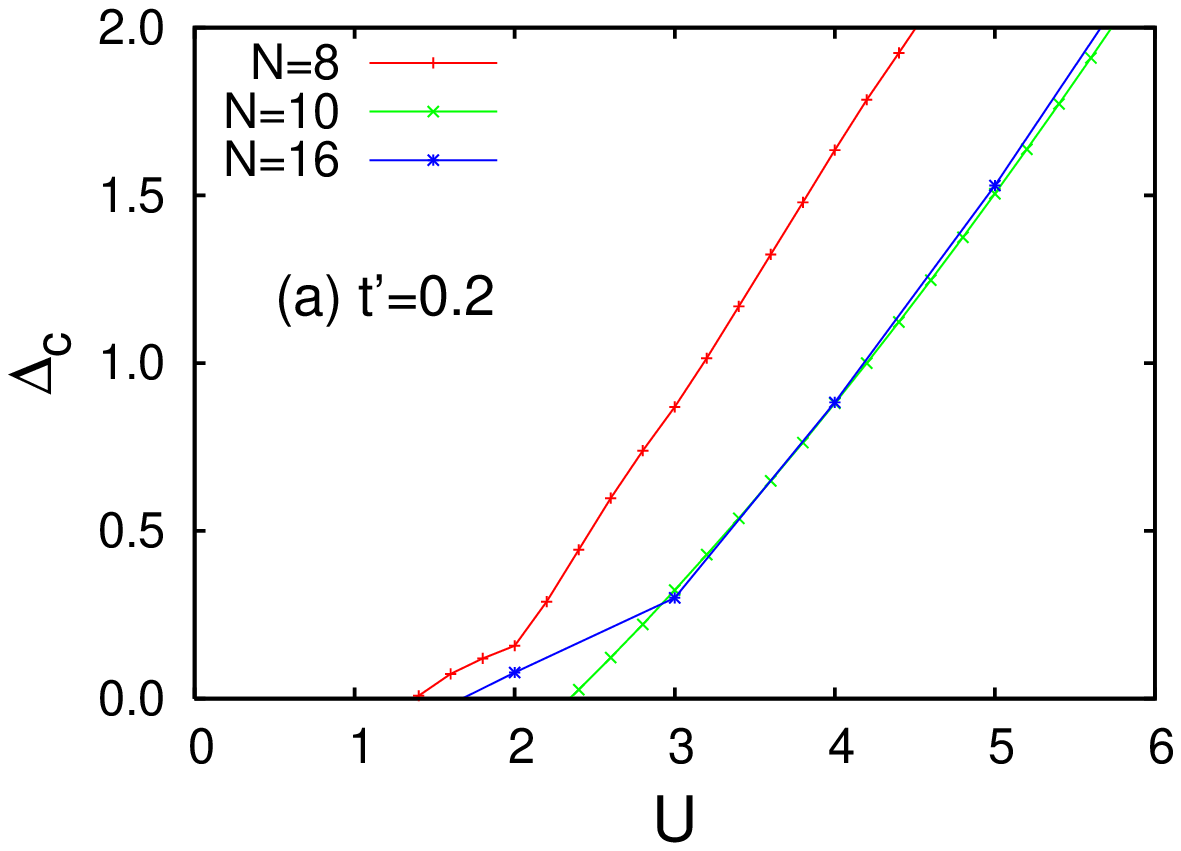}
\includegraphics[scale=0.6]{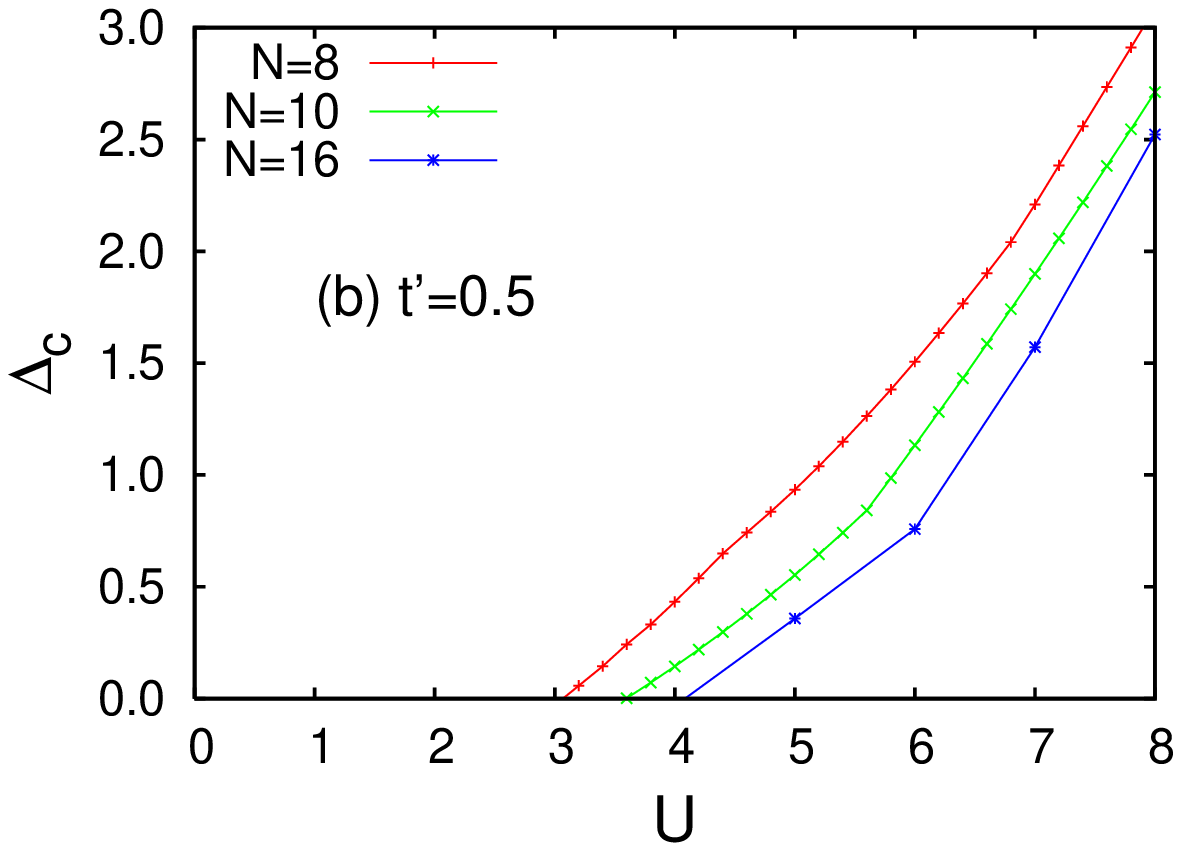}
\includegraphics[scale=0.6]{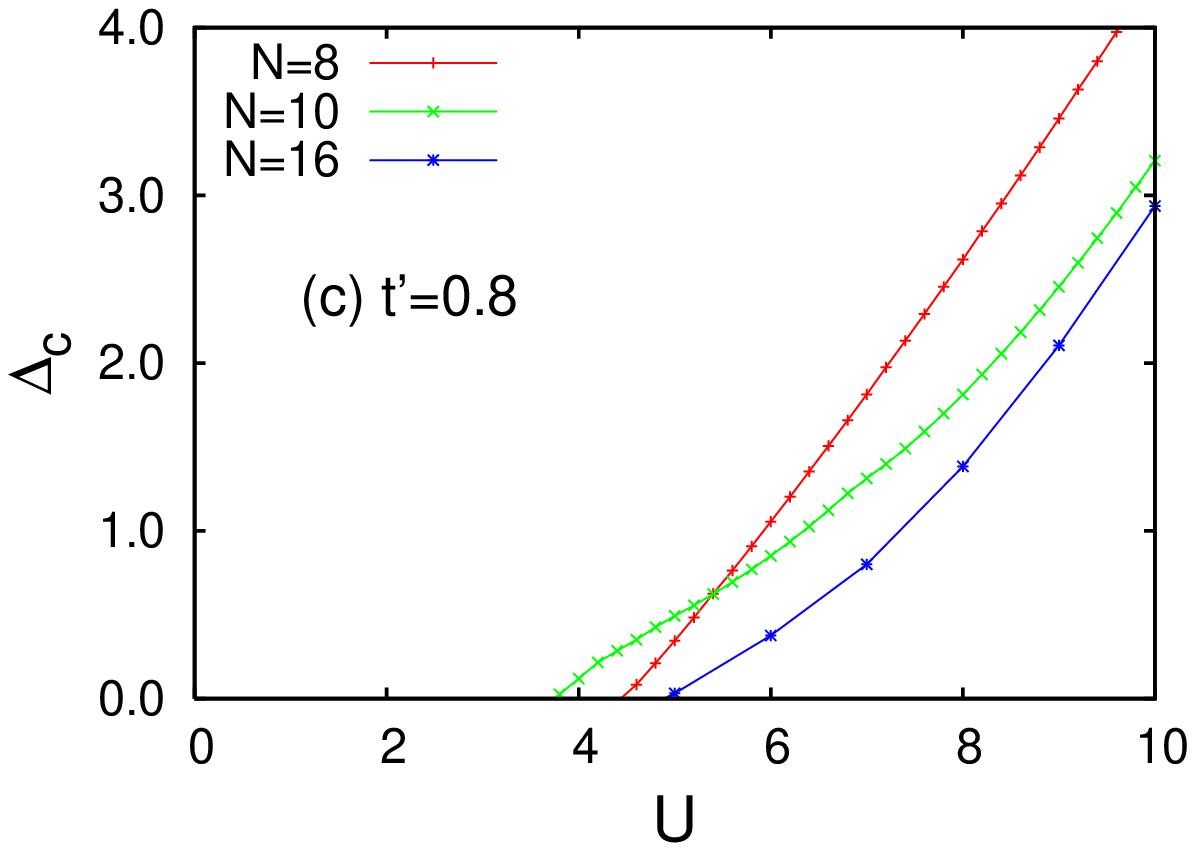}
\caption{
Charge gap at (a) $t'=0.2$, (b) $0.5$ and (c) $0.8$.
}
\label{fig:chargegap}
\end{figure}

Figure \ref{fig:chargegap} shows 
numerical results for the charge gap $\Delta_\mathrm{c}$
defined in eq.\ (\ref{eq:chargegap}).
A finite $t'$ lifts the perfect nesting, and
the charge gap opens up when $U$ is larger than some finite value.
We thus define the second critical value $U_\mathrm{c}^\Delta$
above which we have a nonvanishing charge gap.
Although it is difficult to see systematic size dependence,
we can roughly estimate the critical value as
$U_\mathrm{c}^\Delta = 2 \sim 3$, $3.5 \sim 4.5$ and $4 \sim 6$
for $t'=0.2$, $0.5$ and $0.8$, respectively.

Note that the estimated values of $U_\mathrm{c}^\Delta$ are always 
slightly smaller than $U_\mathrm{c}^D$ obtained from the Drude weight.

\subsection{Double occupancy}
\begin{figure}
\includegraphics[scale=0.6]{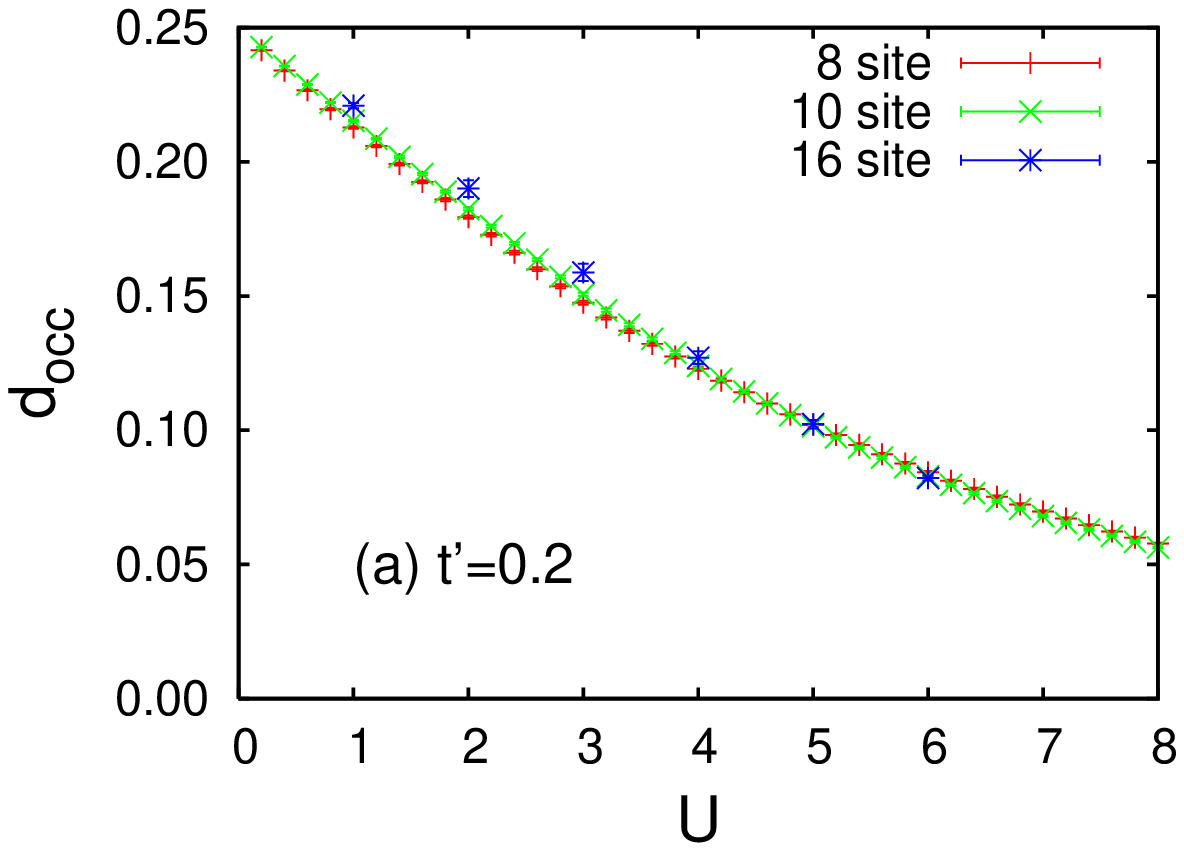}
\includegraphics[scale=0.6]{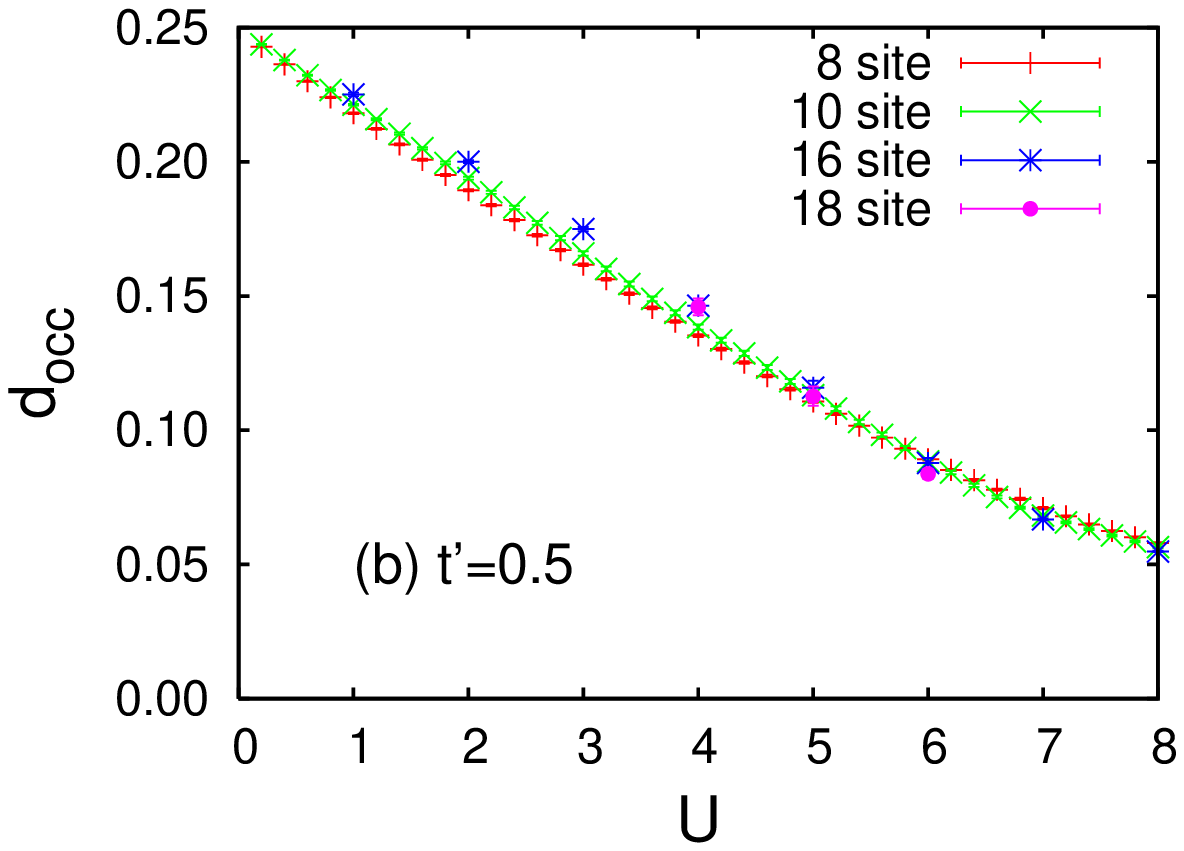}
\includegraphics[scale=0.6]{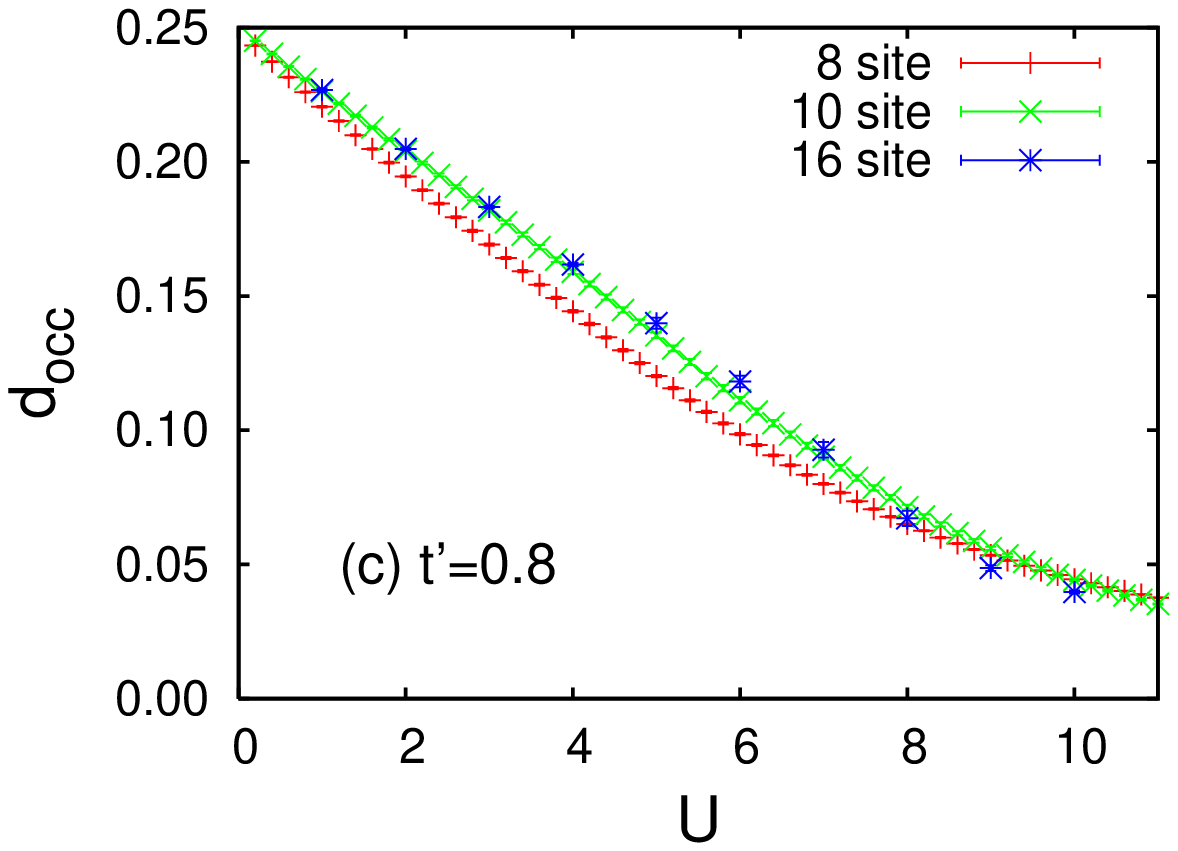}
\caption{Double occupancy at (a) $t'=0.2$, (b) $0.5$ and (c) $0.8$.} 
\label{fig:double}
\end{figure}

Figure \ref{fig:double} shows the double occupancy $d_\mathrm{occ}$
at $t'=0.2$, $0.5$ and $0.8$.
We find that the size dependence of the double occupancy changes 
at some finite $U$.
For example, at $t'=0.5$, 
the double occupancy increases with increasing $N$ for $U<5$ 
while it decreases for $U>5$.
Accordingly, in larger clusters (e.g., $N=16$), the $U$ dependence
has an inflection point. 
We thus define the third critical value $U_\mathrm{c}^d$
from the change in the size dependence.
We estimate $U_\mathrm{c}^d \simeq 4$, $5$ and $7$ 
at $t'=0.2$, $0.5$ and $0.8$, respectively.

The double occupancy is considered to be closely related with
the order parameter of the Mott transition,\cite{kotliar2000}
and has been calculated to determine the transition point
in the previous theoretical studies.
We note that the values of $U_\mathrm{c}^d$ agree well with 
the critical values $U_\mathrm{c}^D$ obtained from the Drude weight
in \S4.1.
The relation 
among $U_{\rm c}^d$, $U_{\rm c}^D$ and $U_{\rm c}^{\Delta}$
will be discussed in \S5 in connection with the phase diagram.

\subsection{Spin structure factor}
In order to find possible magnetic phase transitions,
we have calculated the spin structure
factor, $S(\mib{q})$, defined in eq.\ (\ref{eq:sq}).
For the parameters that we studied, 
$S(\mib{q})$ always shows a peak at $\mib{Q}=(\pi, \pi)$,
except when $t'=0.8$ and $U < 5$ for the 16-site cluster.
Thus we here analyze the staggered component, $S(\mib{Q})$, only.

\begin{figure}
\includegraphics[scale=0.6]{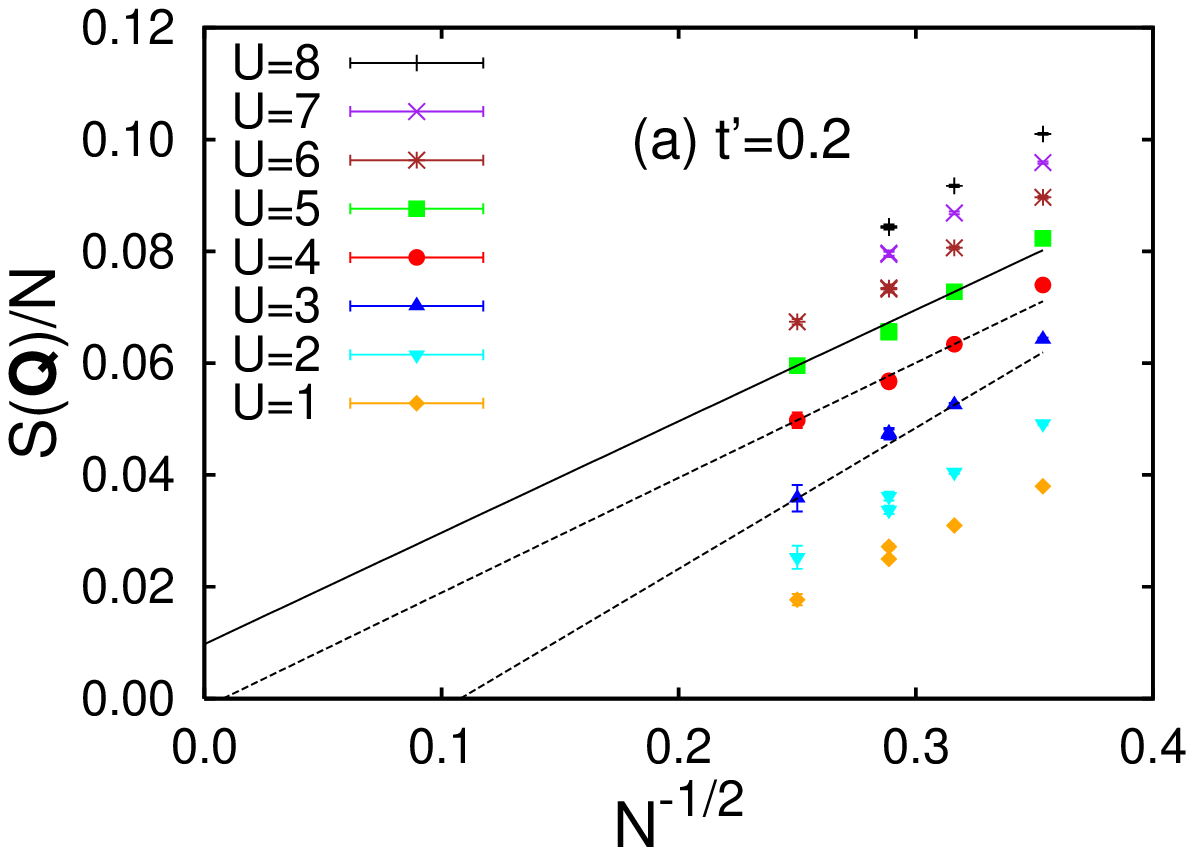}
\includegraphics[scale=0.6]{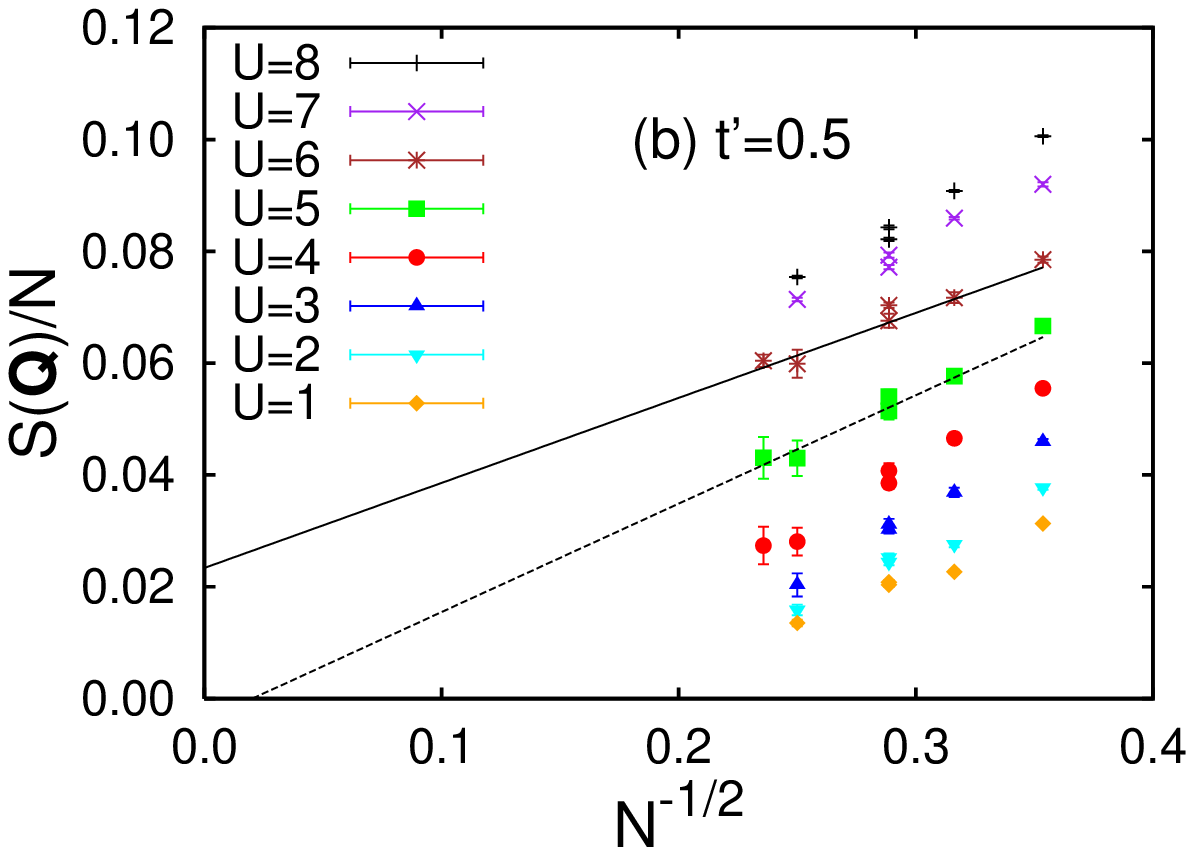}
\includegraphics[scale=0.6]{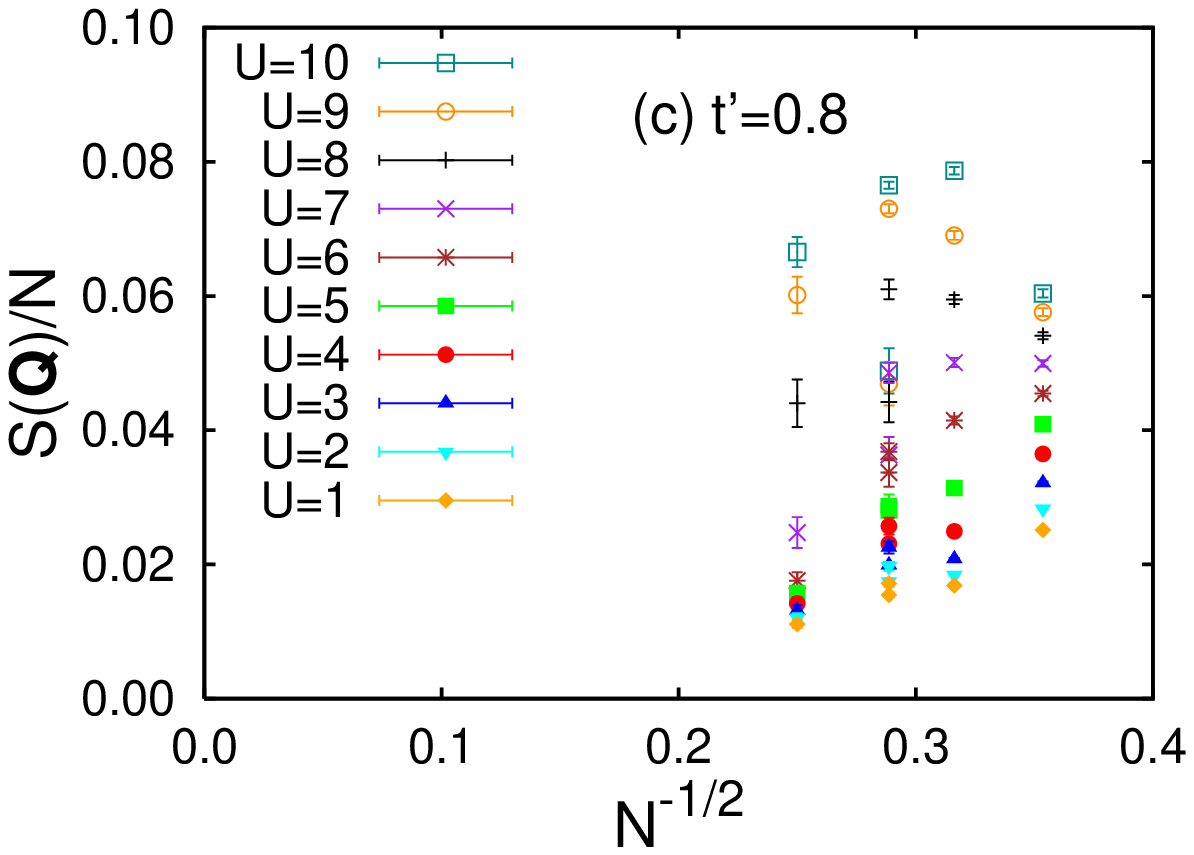}
\caption{
Spin structure factor, $S(\mib{Q})$ with $\mib{Q}=(\pi, \pi)$, 
at (a) $t'=0.2$, (b) $0.5$ and (c) $0.8$.
The lines are 
fits by a linear function of $N^{-1/2}$.}
\label{fig:sq}
\end{figure}

Figure \ref{fig:sq} shows the size dependence of $S(\mib{Q})/N$ 
at $t'=0.2$, $0.5$ and $0.8$.
Here we have also plotted the results for the 12-site clusters. 
Although there are two 12-site clusters with different geometries
as shown in Fig.\ \ref{fig:cluster},
the two results 
agree well, 
except at $t'=0.8$.
For $t'=0.2$ and $0.5$, systematic size scaling is observed, 
whose scaling form is consistent with that expected 
in the staggered ordered phase,
\begin{equation}
\sqrt{S(\mib{Q})/N} - m_\infty \propto N^{-1/2} + \cdots,
\end{equation}
where
$m_\infty = \lim_{N \to \infty} \sqrt{S(\mib{Q})/N}$.
From the value of $U$ where $m_\infty$ becomes positive,
we estimate
the antiferromagnetic transition point to be at
$U_{\rm AF} = 3.5\sim4.5$ and $5\sim6$ for $t'=0.2$ and $0.5$, respectively.
At $t'=0.8$, however, we cannot find any systematic size dependence, 
suggesting that the antiferromagnetic state is not stable 
in this range of $U$.

We note that the size dependences of $S(\mib{Q})$ 
at $t'=0.5$ are consistent with 
the results obtained with the PIRG method
for larger clusters $N = 36 \sim 196$.\cite{morita2002} 
This again confirms that the averaging over the twisted BCs
captures the asymptotic behavior in the limit of $N \to \infty$
from the calculations for small clusters.

\section{Discussion}

In this section we discuss the numerical results presented in \S4
and deduce the ground-state phase diagram.

\subsection{Phase diagram}

In the previous section we have shown our numerical results of
the Drude weight, the charge gap, the double occupancy and
the spin structure factor, which were obtained from
exact diagonalization of finite-size clusters up to 18 sites.
Among them the Drude weight and the charge gap are the quantities 
directly signalling the metal-insulator transition.
The critical values of $U$ estimated from these two quantities,
however, turned out to be slightly different,
$U_{\rm c}^\Delta < U_{\rm c}^D$.
On the other hand, $U_\mathrm{c}^d$ estimated from
the double occupancy was about the same as $U_\mathrm{c}^D$,
i.e., $U_{\rm c}^D \simeq U_{\rm c}^d$.
Naively, these results would imply that,
in the parameter region $U_{\rm c}^\Delta < U < U_{\rm c}^D$,
the charge gap opens but the Drude weight remains finite.
It is, however, difficult to conceive a state with both a finite
charge gap and a finite Drude weight,
and we therefore expect that, in the thermodynamic limit, the Drude weight
should vanish whenever the charge gap opens.
As discussed in the following, we speculate that the seemingly conflicting result
$U_\mathrm{c}^\Delta<U_\mathrm{c}^D$ is due to non-trivial size
dependence of $D$
which cannot be captured by our calculations up to 18 sites.

Here we propose two scenarios for the way in which our finite-size
results approach the thermodynamic limit.
The first scenario is obvious one: as the system size increases,
the difference between $U_{\rm c}^\Delta$ and $U_{\rm c}^D$ is
reduced and eventually vanishes, and we are left with one metal-insulator
transition besides the magnetic transition at a larger
value $U=U_\mathrm{AF}$.
In the other scenario there remain two phase boundaries
corresponding to $U_{\rm c}^\Delta$ and $U_{\rm c}^D$
even in the thermodynamic limit, and
system-size dependence appears mainly in the Drude weight $D$
(in the interval $U_\mathrm{c}^\Delta<U<U_\mathrm{c}^D$)
which decreases to zero,
while the difference between $U_\mathrm{c}^\Delta$ and
$U_\mathrm{c}^D$ is kept finite.

\begin{figure}
\includegraphics[scale=0.6]{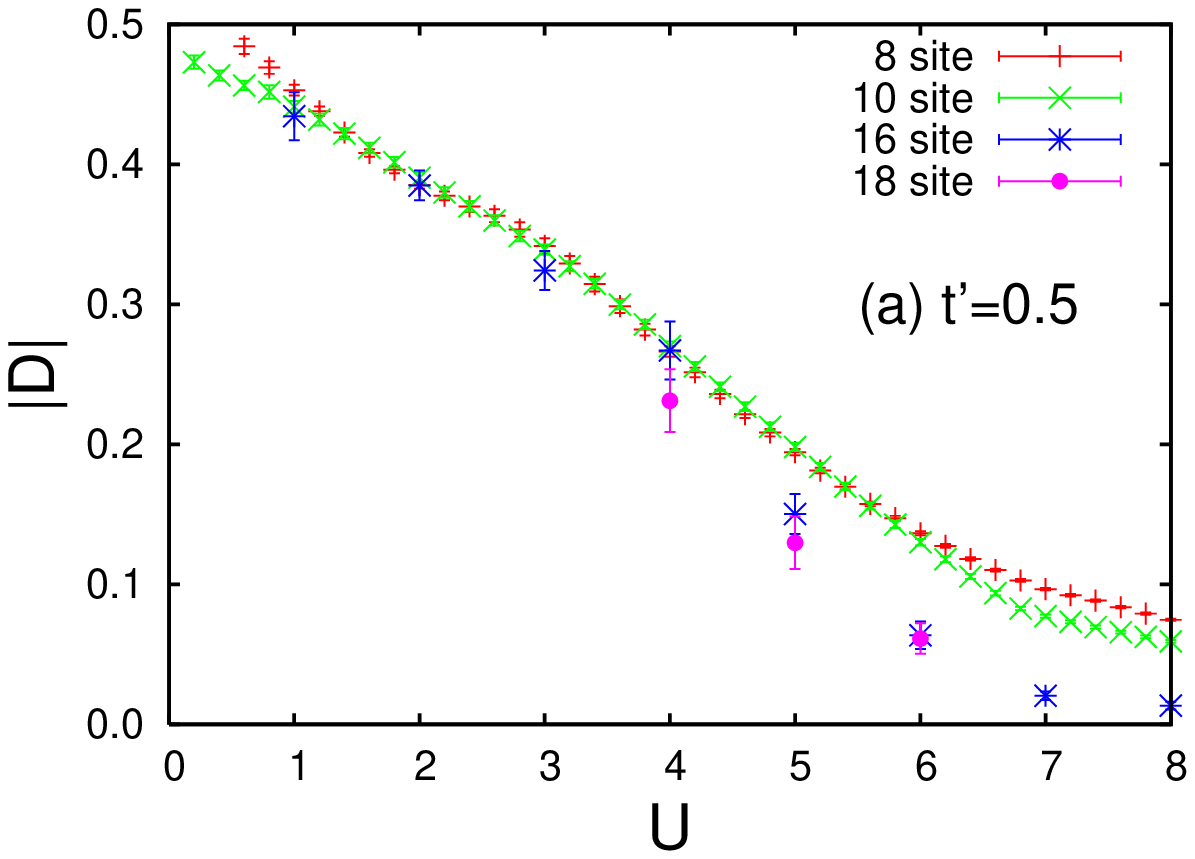}
\includegraphics[scale=0.6]{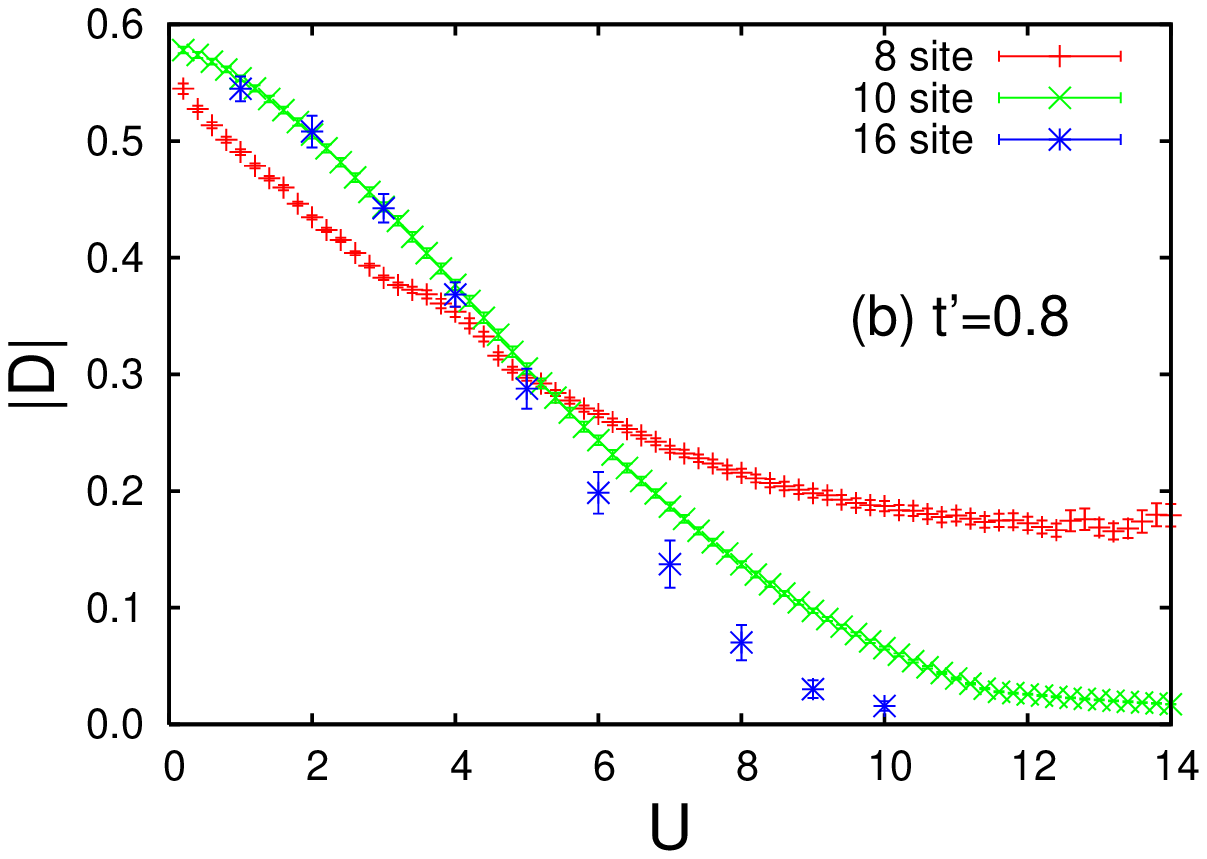}
\caption{
``Absolute value'' of the Drude weight, $|D|$, at (a) $t'=0.5$ and
(b) $t'=0.8$.
}
\label{fig:drude_abs}
\end{figure}

First, let us consider the former scenario
and discuss the size dependence of
the Drude weight and the charge gap.
Concerning the Drude weight, as we mentioned in \S4.1,
the data at $t'=0.8$ clearly show a system-size dependence for $U>6$,
where the Drude weight becomes smaller in larger systems.
Even in the case of $t'=0.5$,
if we look at the data in Fig.\ \ref{fig:drude}(b)
carefully (by neglecting the results for the smallest 8-site cluster), 
we can see smaller but similar system-size dependence for $U>4$,
although the variation of the data for different system sizes are
within the error bars.
We now show that these system-size dependences can be seen
more clearly in the quantity which we denote by $|D|$, i.e.,
the average of the absolute value
of the right-hand side of eq.\ (\ref{eq:drude}) over the twisted BCs.
The numerical results for $|D|$ are shown in Fig.\ \ref{fig:drude_abs}.
At $t'=0.5$, $|D|$ converges on a single curve for $U<4$
while it has systematic size dependence for $U>4$.
A similar size-dependent behavior is clearly seen for $U>5$ at $t'=0.8$.
Although $|D|$ does not have a clear, direct physical meaning,
these apparent system-size dependences of $|D|$
seen in Fig.\ \ref{fig:drude_abs}
suggest that the Drude weight $D$ retain
a size dependence in the same range of $U$.
In other words,
the estimates of $U_{\rm c}^D$ are considered as upper bounds
for the metal-insulator boundary monitored by the Drude weight.
Note that the values of $U$ where the size dependence of $|D|$
becomes evident roughly coincide with 
$U_{\rm c}^\Delta$.

As we have mentioned in \S4.2,
the charge gap does not show systematic
size dependence up to $N=16$.
Incidentally, the PIRG study
has shown that 
the charge gap of a 16-site cluster already gives a good estimate of
that in the thermodynamic limit.\cite{watanabe2003}
If we assume that this is also the case in the present results,
then we may regard $U_{\rm c}^\Delta$ obtained from up to 16-site clusters
as a good estimate for the value in the thermodynamic limit.

These considerations on the system-size dependences
of the Drude weight and the charge gap led us to
the first scenario in which, as $N$ is increased beyond $N=18$,
$U_{\rm c}^D$ decreases 
and finally coincides with $U_{\rm c}^\Delta$ which changes little
with $N$.
In this case the Mott transition takes place at $U_{\rm c}^\Delta$.
The insulating state with $U$ larger than $U_{\rm c}^\Delta$ does not
show the antiferromagnetic order until $U$ reaches
$U_{\rm AF}$ which is substantially larger than $U_{\rm c}^\Delta$.
Although we do not have any further information on the magnetic properties
of the insulating phase in the intermediate-coupling regime,
one possibility is that the insulator is a featureless Mott insulating
state without symmetry breaking, similar to the non-magnetic
insulator found in the PIRG calculations.\cite{morita2002}
Hence, in this scenario, we have three different phases
in the plane of $t'$ and $U$:
a paramagnetic metal for $U < U_{\rm c}^\Delta$,
a non-magnetic insulating state for $U_{\rm c}^\Delta < U < U_{\rm AF}$,
and an antiferromagnetic insulator for $U > U_{\rm AF}$.

Next, we explain the second scenario in which the phase diagram has
two distinct phase boundaries
corresponding to $U_{\rm c}^D$ and $U_{\rm c}^\Delta$.
In this case also, we assume that, in the thermodynamic limit,
the system is insulating for $U > U_{\rm c}^\Delta$ because of the opening
of the charge gap and
because of possibly negligible size dependence of $U_{\rm c}^\Delta$
as discussed above.
One might then wonder how a finite Drude weight observed 
in the region $U_{\rm c}^\Delta < U < U_{\rm c}^D$ diminishes
with increasing system size, while
$U_\mathrm{c}^D$ remains larger than $U_\mathrm{c}^\Delta$.
We speculate that this peculiar behavior of the Drude weight 
is a fingerprint of some unconventional nature of the intermediate phase.
Let us suppose, for example, that we have a bunch of low-energy levels
which are separated by a gap from higher-energy modes.
These low-energy levels are distinguished
by their total momenta $\mib{k}$.
Let us further assume that, as the system size increases,
the band widths of the low-energy modes get narrower,
while their dispersion curves remain to overlap
as in Fig.\ 3(a).
In this case we should observe finite Drude weight which is
decreasing with increasing system size.
When $U_\mathrm{c}^\Delta<U<U_\mathrm{c}^D$, we thus expect to
have an insulating phase with many low-energy modes,
which is reminiscent of the non-magnetic Mott insulator found by
the PIRG study.\cite{morita2002}
Since the twisted BCs are not directly coupled to spin degrees of
freedom, these low-energy modes involve some charge dynamics
and would not be obtained by spin-only models like the Heisenberg model
with multiple-spin exchange interactions.
The remaining problem is what happens at the phase boundary $U_{\rm c}^D$
as $U$ is increased.
A possibility is that there occurs a phase transition to another
insulating state with some conventional symmetry breaking, e.g.,
an insulator with a complicated pattern
of magnetic order or with dimer order.\cite{weihong1999,mizusaki2006}
In such an insulating state, the ground state may consist of 
a single dispersion as depicted in Fig.\ \ref{fig:bcdep}(b), and
the Drude weight becomes zero even in the small size clusters.
Hence, in this second scenario,
we have four different phases:
a paramagnetic metal for $U < U_{\rm c}^\Delta$,
a non-magnetic insulating state for $U_{\rm c}^\Delta < U < U_{\rm c}^D$,
another insulating state for $U_{\rm c}^D < U < U_{\rm AF}$, and
an antiferromagnetic insulator for $U > U_{\rm AF}$.

\begin{figure}
\includegraphics[scale=0.6]{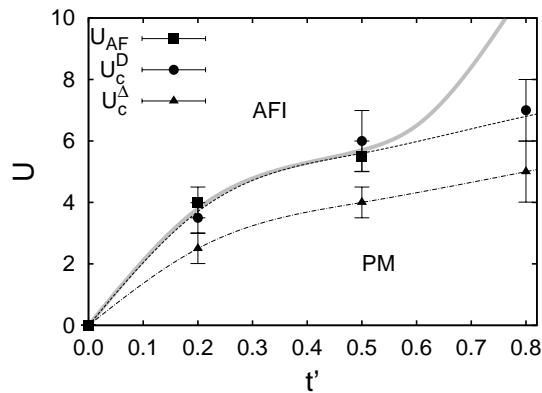}
\caption{
Ground-state phase diagram of the Hubbard model on the anisotropic triangular lattice
in $t'$-$U$ plane.
AFI and PM denote an antiferromagnetic insulating phase and
a paramagnetic metallic phase.
The dash-dotted curve represents the metal-insulator transition
determined by the opening of the charge gap.
The broken curve denotes a possible phase boundary
between different insulating states.
The curves are the guides for eyes.
See the text for details.
}
\label{fig:phasediagram}
\end{figure}

We summarize the possible phase diagram in Fig.\ \ref{fig:phasediagram}
according to the two scenarios explained above.
The grey curve represents the antiferromagnetic transition at $U=U_{\rm AF}$.
The dash-dotted curve corresponds to the opening of charge gap
at $U=U_{\rm c}^\Delta$.
The broken curve denotes a possible phase transition
from an unconventional, non-magnetic insulator to another insulating state
at $U=U_{\rm c}^D$ suggested in the second scenario.

With the numerical results at hand, we cannot conclude 
which scenario is more plausible.
To answer this question requires further studies, such as
calculations on larger clusters,
a through examination of magnetic symmetry breaking 
beyond the simple antiferromagnetic order with ${\mib Q}=(\pi,\pi)$
and of other symmetry breakings (dimerization etc.).

\subsection{Comparison with other theoretical studies}
Here, we discuss the implications of our results
in comparison with other previous theories,
focusing on the nature of the Mott transition.
Comparing the phase diagram in Fig.\ \ref{fig:phasediagram}
with that obtained by the large-scale PIRG calculations,\cite{morita2002} 
we find that the Mott transition line in our phase diagram is
in good agreement with the one determined by the PIRG method.
However, there is an important discrepancy
in the evolution of the double occupancy as a function of $U$.

In our results, as shown in Fig.\ \ref{fig:double}, 
the double occupancy is a smooth decreasing function of $U$,
showing no notable anomaly in the Mott transition at $U_{\rm c}^\Delta$.
On the contrary, in the PIRG results,
it shows a jump at the transition,
indicating the first-order nature of this metal-insulator transition.
Moreover, in the metallic state in the vicinity of the transition,
the double occupancy is not a decreasing function of $U$;
it is almost flat in a certain range of $U$ and
keeps a relatively large value of about 0.2.
This flat regime as well as the jump
is not reproduced in our calculation of the double occupancy.

On the other hand, in the insulating side just above $U_{\rm c}^\Delta$,
the double occupancy shows a good agreement between the two results.
Our results indicate an opening of the charge gap expected there.
Furthermore, as we discussed in \S5.1,
a finite Drude weight showing a small system-size dependence
can be ascribed to some unconventional nature of
the insulating state, as suggested in the PIRG study.\cite{morita2002}
Hence we consider that our present scheme has a potential
to describe the insulating side of the Mott transition.

Thus the most peculiar feature is seen in the metallic side of
the Mott transition.
There the PIRG results indicate that carriers have coherent nature 
even at a moderately large value of $U$.
In fact, the momentum distribution shows a large jump at the Fermi surface
without significant reduction of the renormalization factor.\cite{morita2002}
If we regard the PIRG results as standard results to be reproduced,
then the discrepancy between our results and the PIRG results
implies that the coherent nature of a carrier is not captured 
by our calculations for small size clusters,
even though strong correlations are taken into account properly
in our exact-diagonalization study.
Moreover, the peculiar behavior of the double occupancy
is not reproduced by both the VMC study\cite{watanabe2006}
and the cellular DMFT study.\cite{kyung2006}
Considering the fact that the system size used in the VMC study is much
larger than the one in our study, we suspect that
the discrepancy between the VMC study and the PIRG is indicating that
the many-body effects relevant to the peculiar coherency are not treated properly in the VMC which
employs approximation of a \textit{single} Slater determinant
with a Gutzwiller projection. 
The disagreement between the results of the cellular DMFT and the PIRG
might be ascribed to the problem shared by our study, i.e.,
lack of sufficient information on the spatial fluctuations in larger scale.
Further study is highly desired to clarify the peculiar nature of
the metallic state, and this will help us better understand
the Mott transition.

\section{Summary}
We have studied the ground-state properties of 
the two-dimensional Hubbard model on the anisotropic triangular lattice
by the exact diagonalization method.
In order to reduce the finite-size effects,
we have demonstrated efficiency of averaging over the twisted boundary
conditions for reducing finite-size effects.
Using this technique, we have calculated the Drude weight, the charge gap,
the double occupancy and the spin structure factor, and
determined the the phase boundaries of the metal-insulator transition 
and the antiferromagnetic transition.
An unconventional aspect of the Mott transition of the present system
is illuminated by comparing our results with other previous
theoretical results.

\acknowledgement
We would like to thank
M. Imada,
T. Momoi,
H. Nakano,
T. Oka,
H. Tsunetsugu,
S. Watanabe,
and
Y. Yanase
for useful discussions.
This work was in part supported
by Grant-in-Aid for Scientific Research
(No.\ 16GS0219 and 17071003)
and NAREGI
from Ministry of Education, Culture, Sports,
Science and Technology, Japan.
Numerical calculations are partly performed on
the Hitachi SR11000 of the Supercomputing Division,
Information Technology Center, The University of Tokyo.


\begin{thebibliography}{10}
\bibitem{note2}
We note that our definition of the Drude weight does not
recover the triangular lattice symmetry at $t'=t$.
The definition covers the square lattice symmetry and
is suitable for our study,
since we here focus on the anisotropic cases with $t' < t$
including the square lattice at $t'=0$.
\bibitem{note}
The insulating BC dependence does not necessarily mean that 
the ground state is unique in the thermodynamic limit 
since we only consider the same BC for up and down spins,
which cannot detect the degeneracy of states with different total spin.
In order to study the degeneracy in total spin, which is expected,
for example, in the
antiferromagnetic state, we have to change BC of up and down spins separately,
which is beyond the scope of this paper.
\end{thebibliography}
\end{document}